\RequirePackage[dvipsnames,prologue]{xcolor}
\documentclass[sigconf, screen]{acmart}
\definecolor{green}{RGB}{214,251,225}
\definecolor{blue}{RGB}{213,228,252}
\definecolor{purple}{RGB}{239,227,252}
\definecolor{teal}{RGB}{174,214,224}
\definecolor{mint}{RGB}{184,242,230}
\definecolor{yellow}{RGB}{250,243,221}
\usepackage{csquotes}
\usepackage{listings}
\usepackage{tabularx}
\usepackage{longtable}
\usepackage{ragged2e}

\lstdefinestyle{prompt}{
    basicstyle=\ttfamily\small,
    breaklines=true,
    frame=single,
    frameround=tttt,
    breakindent=0pt,
    columns=fullflexible,
    basewidth=0.95em,
    aboveskip=1em,
    belowskip=1em,
    emphstyle=\textit,
}

\MakeOuterQuote{"}
\AtBeginDocument{%
  \providecommand\BibTeX{{%
    \normalfont B\kern-0.5em{\scshape i\kern-0.25em b}\kern-0.8em\TeX}}}

\copyrightyear{2025}
\acmYear{2025}
\setcopyright{rightsretained}
\setcctype{by}
\acmConference[CHI '25]{CHI Conference on Human Factors in Computing Systems}{April 26-May 1, 2025}{Yokohama, Japan}
\acmBooktitle{CHI Conference on Human Factors in Computing Systems (CHI '25), April 26-May 1, 2025, Yokohama, Japan}\acmDOI{10.1145/3706598.3713397}
\acmISBN{979-8-4007-1394-1/25/04}

\usepackage{enumitem}
\usepackage{multirow}

\newcommand{\rev}[1] {{#1}}

\begin{document}

% \title{Inkspire: Supporting Designers to Prototype Product Designs\\by Sketching with AI}
% \title{Inkspire: Supporting Designers to Prototype Product Designs with AI-assisted Analogies and Sketching}
% \title{Inkspire: Supporting Design Exploration with Generative AI through Analogies and Iterative Sketching}
% \title{Inkspire: Overcoming Design Fixation in Generative AI Image Models using Analogies and Iterative Sketch Inputs}
% \title{Inkspire: Overcoming Design Fixation with Image Generation Models via Analogies and Iterative Sketching}
\title{Inkspire: Supporting Design Exploration with Generative AI through Analogical Sketching}
%% used to denote shared contribution to the research
% \author{Anonymous Author(s)}

\author{David Chuan-En Lin}
\affiliation{%
  \institution{Carnegie Mellon University}
  \streetaddress{5000 Forbes Ave.}
  \city{Pittsburgh, PA}
  \country{USA}
  }
\email{chuanenl@cs.cmu.edu}

\author{Hyeonsu B. Kang}
\affiliation{%
  \institution{Carnegie Mellon University}
  \streetaddress{5000 Forbes Ave.}
  \city{Pittsburgh, PA}
  \country{USA}
  }
\email{hyeonsuk@cs.cmu.edu}

\author{Nikolas Martelaro}
\affiliation{%
  \institution{Carnegie Mellon University}
  \streetaddress{5000 Forbes Ave.}
  \city{Pittsburgh, PA}
  \country{USA}
  }
\email{nikmart@cmu.edu}

\author{Aniket Kittur}
\affiliation{%
  \institution{Carnegie Mellon University}
  \streetaddress{5000 Forbes Ave.}
  \city{Pittsburgh, PA}
  \country{USA}
  }
\email{nkittur@cs.cmu.edu}

\author{Yan-Ying Chen}
\affiliation{%
  \institution{Toyota Research Institute}
  \streetaddress{4440 El Camino Real}
  \city{Los Altos, CA}
  \country{USA}
  }
\email{yan-ying.chen@tri.global}

\author{Matthew K. Hong}
\affiliation{%
  \institution{Toyota Research Institute}
  \streetaddress{4440 El Camino Real}
  \city{Los Altos, CA}
  \country{USA}
  }
\email{matt.hong@tri.global}

\renewcommand{\shortauthors}{David Chuan-En Lin, Hyeonsu B. Kang, Nikolas Martelaro, Aniket Kittur, Yan-Ying Chen, Matthew K. Hong}

\begin{abstract}
With recent advancements in the capabilities of Text-to-Image (T2I) AI models, product designers have begun experimenting with them in their work. However, T2I models struggle to interpret abstract language and the current user experience of T2I tools can induce design fixation rather than a more iterative, exploratory process. To address these challenges, we developed Inkspire, a sketch-driven tool that supports designers in prototyping product design concepts with analogical inspirations and a complete sketch-to-design-to-sketch feedback loop. To inform the design of Inkspire, we conducted an exchange session with designers and distilled design goals for improving T2I interactions. In a within-subjects study comparing Inkspire to ControlNet, we found that Inkspire supported designers with more inspiration and exploration of design ideas, and improved aspects of the co-creative process by allowing designers to effectively grasp the current state of the AI to guide it towards novel design intentions.
\end{abstract}


% \ccsdesc[500]{Applied computing~Arts and humanities}


% \textwidth

\begin{teaserfigure}
  \centering
  \includegraphics[width=12cm]{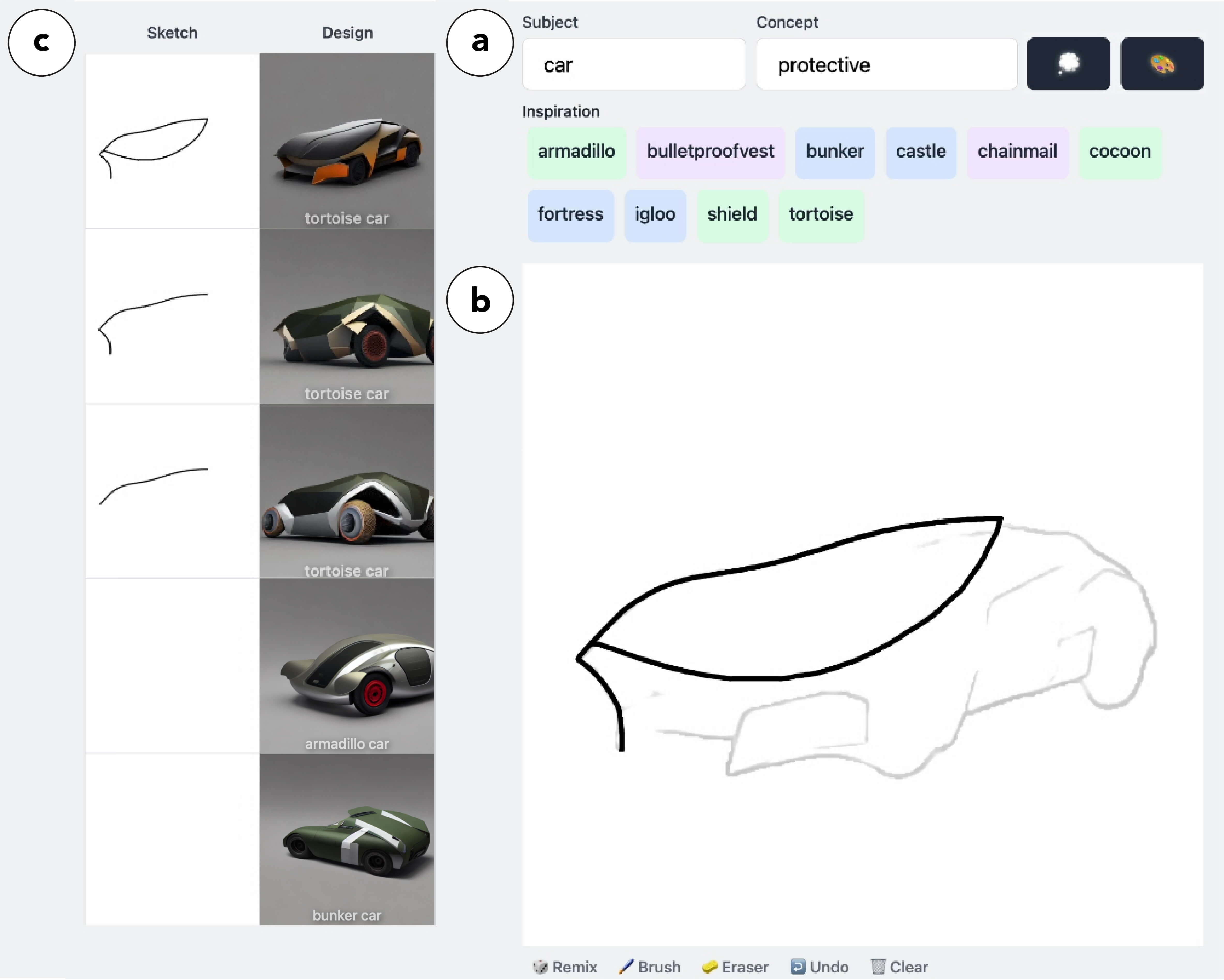}
  \caption{The Inkspire interface. The designer may use the Analogical Panel (a) to ideate analogical inspirations for abstract concepts (e.g., "protective car" $\rightarrow$ "tortoise car"). The designer may sketch on the Sketching Panel (b) to iteratively guide AI design generations. For each iteration, we display a sketch scaffolding under the canvas. This scaffolding is created through abstracting AI designs into lower fidelity. Finally, the designer may view the history of iterations on the Evolution Panel (c).}
  \label{fig:teaser}
\end{teaserfigure}

% Inkspire facilitates an iterative sketching process between the designer and AI. The designer may start with a product category (e.g., chair), an abstract concept to design around (e.g., fluidity), and a simple line stroke. Inkspire helps generate a set of visually-concrete analogies to ground the abstract concept (e.g., fluidity \rightarrow river, staircase, ribbon). Guided by the stroke and analogy inspiration, the AI generates a design. Then, the AI converts the design back into a sketch-style scaffolding, which underlays the designer's canvas. Using the scaffolding, the designer can draw inspiration from the AI's creations and further iterate upon them, receiving a new design generation with each new pen stroke.

\maketitle
\pagestyle{plain}

\section{Introduction}

We have seen significant progress in the capabilities of text-to-image (T2I) models, many of which are now able to generate realistic images using text \cite{betker2023improving}.
These models not only accelerate the process of converting thoughts into visuals but can also potentially create serendipitous inspirations for users \cite{rooij2024expecting}.
Recent research has also opened up new possibilities for translating one image representation into another, such as transforming a sketched drawing into detailed designs \cite{zhang2023adding}. Consequently, many designers have begun embracing the use of T2I models to enhance their creative work.

However, despite the proposed benefits T2I models, integrating them into the designer's creative workflow can be challenging. In particular, recent works have observed that designers experience \textit{a high level of fixation} \cite{jansson1991design} when using generative AI \cite{Wadinambiarachchi2024effects} leading them to explore fewer novel ideas than may otherwise be beneficial for innovation.
For instance, a designer might write a text prompt and hit generate. Upon viewing the result, they might adjust a few words in their prompt and hit generate again \cite{chang2023prompt}---repeating this process akin to using a slot machine \cite{ai-slot-machine}. Their final prompt tends to be conceptually similar to their original one, often with small, incremental modifications aimed at getting the AI to create their \textit{initial design intention}, instead of exploring new diverse design spaces \cite{Wadinambiarachchi2024effects}.
As found in prior work, unsupported text-only prompting can be a limiting interface for generating good outputs from GenAI \cite{zamfirescu2023johnny}.

To understand how professional designers might experience these challenges with GenAI in their own processes, we conducted a comprehensive day-long exchange session with a team of professional designers from a large automotive company.
The team outlined their design process, from conceptualizing a product design sketch based on a specific guiding theme to presenting final designs to stakeholders. 
From our discussions with the designers, we identified three key challenges in their use of GenAI:

\begin{enumerate}[label=C\arabic*]
\item Designing through text prompts feels \textit{unnatural} as compared to their traditional sketching and ideation process.
\item T2I models struggle with generating inspiring designs from \textit{abstract concepts} (e.g., "protective" car), a technique the team uses to create more novel designs.
\item It is difficult to directly build on generated designs---they appear \textit{"too complete"}, potentially leading to fixation.
\end{enumerate}

To help designers working with GenAI avoid design fixation, we might take inspiration from design research working to break designer fixation. 
Two common strategies in line with what we learned from the professional design team include analogy-driven design \cite{goel1997design,linsey2012analogy,jiang2022data}, where concepts from outside the domain area are used to foster inspiration in the domain area (addressing C2) and providing multimodal interfaces that work visually and with lower-fidelity assets (addressing C1 \& C3).

Several recent works in HCI have focused on developing new interfaces for GenAI to help designers explore a larger design space, including interactive prompting which aims help people who often have limited prompting ability \cite{brade2023promptify}, multimodal search \cite{son2024genquery}, and visually navigating a 2D latent space interface \cite{davis2024fashioning}. 
Nonetheless, being able to help designers move beyond one thread of thinking and  explore a wider design space while still taking inspiration and \textit{building on} generated designs remains a significant limitation preventing GenAI to be fully integrated into designers' work. While the ML community offers methods to increase editing freedom of generated images, including local inpainting \cite{saharia2022palette} and instruction-based image editing \cite{brooks2023instructpix2pix}, these methods remove designers away from their natural rhythm of interaction with ideas.

In this paper, we introduce a workflow of continuous exploration with T2I models, encouraging designers to adopt a mindset that facilitates more iterative and exploratory design generation.
Specifically, we built Inkspire, a proof-of-concept tool that provides a more familiar interaction built around iterative sketching and by leveraging the concept of analogical design to facilitate inspiration around design ideas.
Our tool integrates analogical inspiration to promote concept-level ideation from abstract themes, allowing designers to recognize creative possibilities without needing to come up with them and write prompts manually. 
This reduces cognitive friction, enabling fluid exploration of new ideas \cite{budiu2014memory}.
Additionally, inspired by prior works on drawing assistance systems that convert photographs into sketches to teach people how to draw \cite{lee2011shadowdraw}, we introduce a new mechanism that converts high-fidelity AI designs into high-quality but low-resolution sketch scaffolds, directly underlaid on the designer's canvas.
The scaffolding provides a transparent view into the current state of the AI and helps designers build on AI designs without being overly fixated on photorealistic renders.
Finally, we enable a new design generation every time a new pen stroke is drawn (with near real-time performance), encouraging designers to consider small but meaningful changes in form and refinement with each pen stroke, thereby creating many more opportunities for designers to explore new directions.
All of these components are designed to be seamlessly integrated into the sketch process familiar to designers.

To understand how designers use Inkspire, we invited both professional designers and novice users to design everyday products. We also asked them to use a baseline condition of a state-of-the-art ControlNet \cite{zhang2023adding}, which allows designers to use sketches and user-written prompts to guide generative design, but does not provide our proposed analogy inspirations or convert designs into lower resolution sketches, and requires the user to choose when they generate explicitly. 
The results show that users rated Inkspire as providing significantly more inspiration and exploration over baseline ControlNet. 
The interaction that users had with Inkspire was drastically different than using a baseline ControlNet---designers using Inkspire generated many more concept sketches and showed a process that appears more co-creative, whereas when using ControlNet they focused on manual sketching and refining prompts before handing off to the generative system.
Inkspire enabled designers to more effectively co-create designs with T2I models with significantly increased partnership, controllability, communication, and sense of attribution over their creations.
% Designers were able to focus on iterative sketching rather than iterative prompting, focusing more on "getting the right design" instead of "getting the design right" \cite{buxton2010sketching}.
% In addition, Inkspire facilitated a divergent exploration process by allowing designers to generate multiple inspirations and remix designs to explore a broader design space. 
% It also facilitated a convergent refinement process by providing structure for designers to iterate on AI-generated designers. 
% Mechanisms such as scaffolding and the ability to view an overview of design iterations helped designers better understand AI's current state and steer the AI toward their design intentions.
This research thus contributes:
\begin{itemize}
\item{\textbf{Inkspire}, a proof-of-concept, iterative sketching tool that helps ground abstract concepts into analogies and converts AI-generated images into abstracted sketch scaffolds to support fluid design iteration and avoid fixation.}
\item{A \textbf{within-subjects study} showing that Inkspire embodies a more iterative and exploratory workflow with T2I models, with designers rating Inkspire with significantly higher inspiration, exploration, and attributes of co-creation.}
\end{itemize}

\section{Related Work}

This work builds on prior research in human-AI co-creativity for helping people generate ideas \cite{lin2023beyond}.
Specifically, we build upon prior works aiming to help designers create new visual concepts through text to image models and visual inputs \cite{liu2023generative,hou2024c2ideas,lin2023jigsaw}.
We review how generative AI systems have been shown to increase idea generation, but counterintuitively can lead to more design fixation.
We then review possible solutions for overcoming design fixation including analogy-driven design, computational sketching tools, and reducing the fidelity of generated images.

\subsection{Generative AI and Design Fixation}
One of the proposed benefits of generative AI is to help designers avoid \textit{design fixation}, where designers remain stuck in a single line of thinking and a limited set of ideas, thus limiting the conceptual novelty of ideas and potential for innovation \cite{jansson1991design}.
While many of the works suggest that generative AI can help people generate more ideas, potentially helping them to move into new conceptual spaces, recent experimental work has found contradictory results on generative AI's impact on design fixation.
DesignAID \cite{cai2024DesignAID} leverages large language models and image generators to help people explore a large, diverse space of ideas and was found to provide more inspiration than search-based tools.
However, people rated the generative AI ideas as less valuable, with the paper authors suggesting that the ideas may not have been diverse enough, not well matched to the problem, or just not enough to break people design fixation.
Bordas et al.~\cite{bordas2024switching} find that people using ChatGPT 3.5 to help generate ideas to protect and egg falling from 10 meters increased their idea generation but remained overly fixated on specific solutions.
Wadinambiarachchi et al. ~\cite{Wadinambiarachchi2024effects} find that participants using a text-to-image generator to create a new chatbot avatar concept had significantly higher design fixation with fewer ideas, less variety, and less originality when using a text-to-image generator as compared to using an image search engine or coming up with ideas unassisted.
The authors suggest that people's limited prompting ability, a known issue when lay people try to use LLMs \cite{zamfirescu2023johnny}, led them to simply copy keywords from the design brief, limiting the language used to generate the images.
Davis et al.~\cite{davis2024fashioning} and Zhang et al. \cite{zhang2023generative} report similar findings, where designers working with text-only image generators showed limited creative exploration, again due to people's limited abilities with text prompting.

To overcome the fixating issues of text-based generative systems, Davis et al.~\cite{davis2024fashioning} developed a generative system taking example images of dresses and provided graphical buttons and sliders to generate more realistic vs. creative ideas and alter shape, color and texture. 
This visual generative system was preferred by designers and lead to more creative idea explorations.
However, the high-fidelity images presented by many image generation tools, which Wadinambiarachchi et al. ~\cite{Wadinambiarachchi2024effects} suggest could lead to more fixation based on prior design research showing that high-fidelity images lead to more fixation over rougher sketches \cite{cardoso2009design, cheng2014new}.

In our work, we explore how to overcome the fixating issues that many generative AI systems have today.
First, we look to help designers explore more novel ideas by providing interfaces that overcome their limited prompting abilities and help them generate more novel ideas.
For this, we look to analogy-driven design as a potential solution.
Second, we move away from text-only generation and provide more visual interfaces for design exploration (C1), helping better match how designers come up with ideas.
We explore how iterative, computational sketching can be used as an visual input to image generation.
Third, we break from showing only \textit{high-fidelity} images which could lead to design fixation and explore how AI generated images can be altered and presented to designers in \textit{lower resolution} forms to see if such representations may scaffold new ideas.

\subsection{Analogy-Driven Design}
Analogy-driven design, or design-by-analogy, is an approach to drawing inspiration from a known domain, including concepts and products, to find novel solutions to a target domain. An abundance of text-based research systems and prototypes such as DANE \cite{DANE}, Idea-Inspire 4.0 \cite{siddharth2018evaluating}, and BioTRIZ \cite{BioTRIZ} have been developed to support design ideation by offering different design-by-analogy capabilities that include retrieval and mapping of analogies to a target problem based on inferred similarities. 

Recent proliferation of text and image data repositories combined with advances in vision and language models gave rise to new multimodal approaches that expand analogy-driven design to the visual domain \cite{jiang2022data}. For instance, Kwon et al. developed an approach that leverages visual similarity to discover visual analogies for generating new ideas \cite{kwon2019visual}. In addition, Zhang and Jin proposed an unsupervised deep learning model, Sketch-pix2seq, to extract shape features from Quickdraw sketches, creating a latent space that enables defining visual similarities and searching for analogical sketches \cite{zhang2020unsupervised}. Jiang et al. developed a CNN-based model to create feature vectors representing patent images, which combine visual and technological information to enhance visual stimuli retrieval \cite{jiang2021deriving}. While these models present promising methods to support image-based analogy-driven design, the specialized nature of these models reduces their practical appeal to designers seeking exposure to out-of-distribution inspirations.

Large pre-trained models provide exciting opportunities to support domain-agnostic analogy based image retrieval. While T2I models alone struggle to generate images from an abstract concept such as "mystery" \cite{khaliq2024comparison}, machine learning research has demonstrated the use of LLMs to convert abstract concepts into semantically meaningful physical representations, thereby streamlining the process for downstream T2I generation tasks \cite{liao2024text,fan2024prompt,wu2024imagine}. For instance, Fan et al. introduced an approach that extends an abstract concept such as peace with concrete objects (e.g., white doves, olive branches), then rewriting the original prompt to incorporate the objects in a scene \cite{fan2024prompt}. Liao et al. proposed the Text-to-Image generation for Abstract Concepts (TIAC) framework that builds on a three-layer artwork theory to clarify the \textit{intent} of the abstract concept with a detailed definition, then transforming it into semantically-related physical \textit{objects} and concept-dependent \textit{form} \cite{liao2024text}. However, much of this work focused on scene illustration with multiple objects. Moreover, techniques that rely on automated prompt enrichment reduce the steerability of T2I models. In Inkspire, we apply a similar process through the use of LLMs and and analogical reasoning to convert an abstract concept into individual physical objects from across multiple domains. We investigate how providing users with a menu of analogical concepts can help the user rapidly generate diverse analogical-grounded designs with T2I models, to overcome the current challenges of creating inspiring designs from abstract design prompts (C2) in ways that also offer them more control over design space exploration.

%\textbf{TO DO} VisiBlends: A Flexible Workflow for Visual Blends \cite{chilton2019visiblends}. PopBlends: Strategies for Conceptual Blending with Large Language Models \cite{wang2023popblends}. CreativeConnect: Supporting Reference Recombination for Graphic Design Ideation with Generative AI \cite{choi2024creative}. I spy a metaphor: Large language models and diffusion models co-create visual metaphors \cite{chakrabarty2023spy}
%Khaliq et al. examined the capabilities of four state-of-art T2I models in generating images with abstract concepts in the form of verb-object noun pairs (e.g., "remain mystery") and found their performance to be inconsistent

\subsection{Guided Sketching}
Research has explored various computational techniques that provide sketch guidance to aid in several use cases including skill building, serving as reference points, and encouraging creative exploration.

Many existing systems have explored guided sketching tools that help novices learn how to sketch. ShadowDraw \cite{lee2011shadowdraw} offers real-time shadow-based feedback, while systems such as The Drawing Assistant \cite{iarussi2013drawing} and Painting with Bob \cite{benedetti2014painting} focus on translating photographs to sketches. Several works have explored crowdsourcing-based approaches. Systems like Limpaecher et al. \cite{limpaecher2013real} and Sketchy \cite{sangkloy2016sketchy} leverage collective human knowledge to provide guidance. These systems excel at teaching specific drawing techniques but are less suited for open-ended creative exploration. Similarly, portrait-specific systems such as DualFace \cite{huang2022dualface}, which employs a two-stage drawing guidance for freehand portraits, and PortraitSketch \cite{xie2014portraitsketch}, which provides face sketching assistance, demonstrate the value of domain-specific guidance but are constrained to a narrow use case.

Most relevant to Inkspire is Creative Sketching Partner \cite{davis2019creative}, which retrieves sketches that are visually and conceptually similar to the user's sketches as a means to stimulate exploration and inspire new designs. This system demonstrates the potential of computational guided sketching systems to inspire new designs. However, it has the limitation of relying on existing sketch databases, whereas in this work, we leverage the generative capabilities of AI.

In our work, we draw inspiration from empirical research by Williford et al. \cite{williford2023exploring}, whose analysis of 240 concept sketches revealed that ambiguous sketch underlays could reduce fixation on conventional forms and promote divergent thinking, guided sketching techniques such as ShadowDraw \cite{lee2011shadowdraw}, and existing design practices of sketch scaffolding \cite{youtube}. With Inkspire, we propose a novel computational pipeline for converting GenAI designs into abstracted, yet high-quality sketching scaffolds that underlay the user's canvas. To the best of our knowledge, this work is the first to support the full closed-loop-cycle of sketching GenAI designs and abstracting GenAI designs into sketches. Through this technique, we address the challenge of building on GenAI images that are "too complete" (C3) by drawing designers' attention away from the high-fidelity generated image and inspire designers to iterate on top of the silhouette of GenAI designs with low friction.

\section{Formative Session with Design Professionals}

To understand how professional designers use Text-to-Image (T2I) models in their work, we conducted a day-long exchange session with a team of seven professional product designers from a large automotive company.
The designers in our exchange session work at a top 5 automotive manufacturing company and 
cover multiple disciplines of training spanning a wide range of roles in the company, including creative director, modeling lead, interior designer, exterior designer, artistic creator, conceptual lead creator, UX/UI and strategy.
There is significant cross-collaboration across departments globally and with other companies in the industry.
The designers also have experience in using T2I tools such as Midjourney \cite{midjourney} and Vizcom \cite{vizcom}.
The designers showed our team their design processes in presentations with specific examples from their past work. This included sketches, concept boards, and various documentations of collaborative meetings for a wide range of mobility concepts. The designers also showed their process of using current T2I tools, and we discussed new ideas on how to design tools to support them.
From our interaction with the designers, we identified key challenges they face when using T2I models and summarized them into three design goals to inform the development of Inkspire.

\subsection{Design Goals}

\subsubsection{Design Goal 1. Sketching as a Natural Method of Interaction}

Designers emphasized that prompting is an unnatural approach to designing. They expressed difficulty in effectively conveying design ideas through language \cite{zamfirescu2023johnny}. They often felt constrained by the need to craft comprehensive prompts, which limited their ability to explore a wider range of ideas. In contrast, designers expressed a preference for approaching design tasks through \textit{sketching} \cite{buxton2010sketching}, often beginning with just a simple line or silhouette. Therefore, our first design goal is to allow designers to interact with AI via sketching as a natural method of interaction. We aim to support designers in starting with simple abstract lines and assist them in progressing towards complete sketches.

\subsubsection{Design Goal 2. Visually-Concrete Inspirations}

Designers mentioned that design briefs are typically inherently abstract, for example, "design a vehicle that conveys a sense of protectiveness". However, they found T2I models to generally produce poor and generic results when prompted with such abstract terms \cite{xie2023prompt}. Even when resorting to prompt engineering tricks, they find it challenging to visualize abstract concepts in concrete forms. Therefore, our second design goal is to assist designers in visualizing abstract design themes through visually-concrete inspirations. Inspired by the way designers may draw inspiration from nature \cite{deldin2013asknature}, we aim to recommend analogical inspirations \cite{kang2023biospark} to designers and make it easy for them to quickly visualize a variety of diverse inspirations.

\subsubsection{Design Goal 3. Complete the Feedback Loop}

Designers expressed difficulty in iterating on AI-generated designs and they described the process of using T2I models as a one-way process. They feel that the generated designs look "too complete", making it difficult for them to envision new ways to build on them. Often, designers find themselves in a position where they can either choose to use a design or discard it entirely. Therefore, we aim to bridge the gap in the feedback loop -- while T2I models transform ideas into images, our goal is to \textit{transform images back into abstractions (i.e., sketches), to allow designers to continue the iteration process.}

% \begin{figure*}[tbp]
%   \centering
%   \includegraphics[width=9.5cm]{figures/interface.jpg}
%   \caption{The Inkspire interface consists of (a) the Analogical Panel, (b) the Sketching Panel, and (c) the Evolution Panel.}
%   \Description{The Inkspire interface consists of (a) the Analogical Panel, (b) the Sketching Panel, and (c) the Evolution Panel.}
%   \label{fig:interface}
% \end{figure*}

\section{Inkspire}
\label{section:implementation}
We first illustrate how a user would use Inkspire through a concept car design example.
We then describe the technical implementation of Inkspire, which consists of two primary components: Sketch2Design (generating AI designs from sketches and analogies) and Design2Sketch (converting AI designs into lower fidelity sketch scaffolding).

\subsection{System Walkthrough}

DeLorean is an automotive designer tasked with creating a concept car design that embodies a sense of "protectiveness" (Figure \ref{fig:teaser}).

\subsubsection{Ideating a Design Concept}

To begin, DeLorean uses the \textit{Analogy Panel} (Figure \ref{fig:teaser}a) to generate visually-concrete inspirations for the abstract concept (Design Goal 2). He uses "car" as the subject and "protective" as the abstract concept, then clicks on the inspiration button. He is presented with a selection of inspirations, color-coded by categories: \colorbox{green}{nature}, \colorbox{purple}{fashion}, and \colorbox{blue}{architecture}. DeLorean experiments with several inspirations (such as bunker, armadillo, tortoise) by clicking on them. For each inspiration he selects, the AI generates a design in the \textit{Evolution Panel} (Figure \ref{fig:teaser}c). He can manually edit inspirations in the concept box and manually click on the generate button. DeLorean decides to use \colorbox{green}{tortoise} as his inspiration.

\subsubsection{Iterating on Designs through Sketching}

DeLorean now iteratively guides the AI through sketching in the \textit{Sketching Panel} (Figure \ref{fig:teaser}b) (Design Goal 1). He starts off with a simple silhouette line, and the AI generates an initial design in the Evolution Panel. In the Sketching Panel, DeLorean sees a \textit{scaffolding} abstracted from the initial AI design (Design Goal 3). This allows him to take inspiration from the AI designs without being overly fixated on photorealistic renders. He is drawn to the bold curve of the windshield area shown in the scaffolding and loosely traces this part to add to his sketch. The AI then generates a new design and scaffolding.

DeLorean repeats this back-and-forth process with the AI (Design Goal 1), continuing to iterate on his designs through sketching until he achieves a design that satisfies him. To navigate between iterations, restart a sketch, or refine parts of the sketch, he uses the \textit{Undo}, \textit{Clear}, and \textit{Eraser} tools. To explore design variations using the same input (the same inspiration and sketches), he uses the \textit{Remix} tool.

\begin{figure*}[tbp]
  \centering
  \includegraphics[width=14cm]{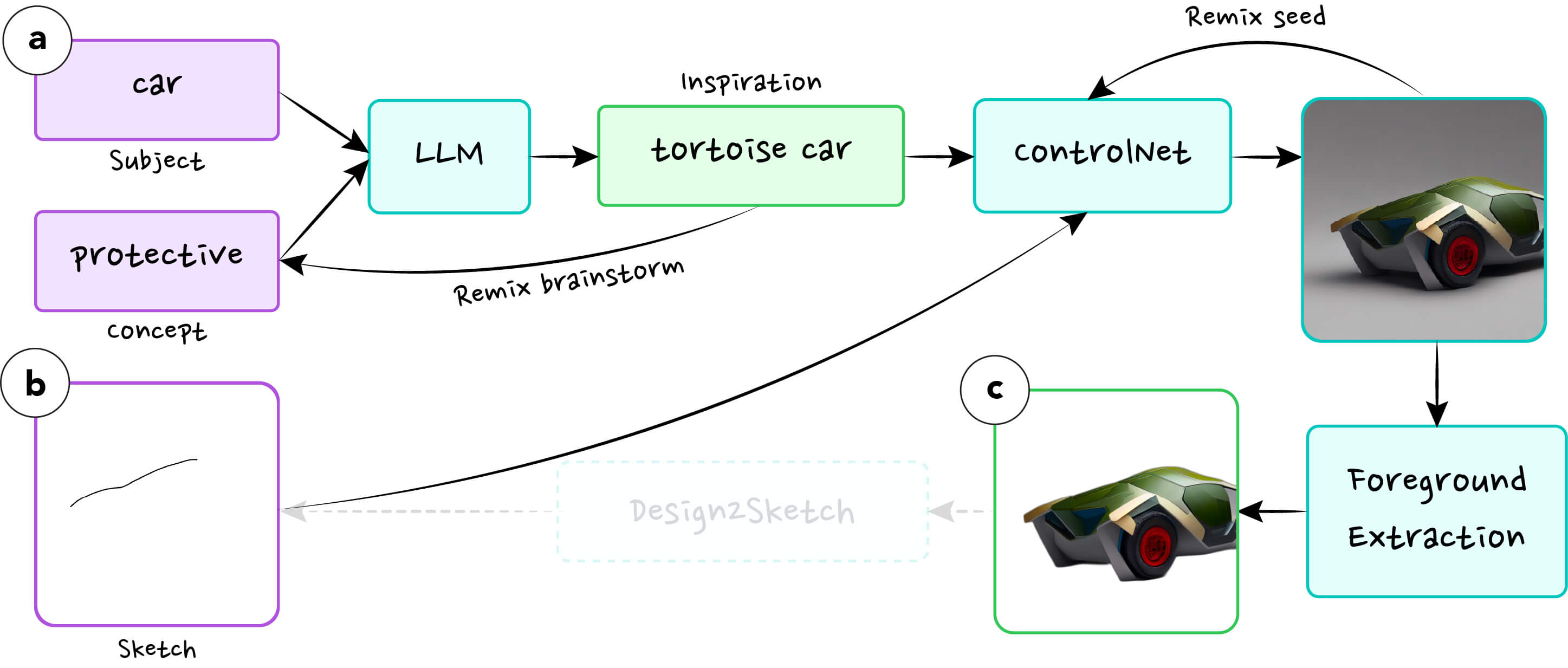}
  \vspace{-1em}
  \caption{Sketch2Design pipeline, including (a) inspiration generation with LLMs, (b) sketch-guided design generation, and (c) foreground extraction.}
  \Description{Sketch2Design pipeline, including inspiration generation with LLMs, sketch-guided design generation, and foreground extraction.}
  \label{fig:sketch2design}
\end{figure*}

\subsection{Sketch2Design}

The \textit{Sketch2Design} component helps users brainstorm design concepts and generate product designs through sketching (Figure \ref{fig:sketch2design}). First, the user specifies the subject that they are designing for (e.g., car) and an initial abstract concept (e.g., protective). 
To brainstorm more concrete design ideas for the abstract concept (Design Goal 2), we leverage Large Language Models (LLMs) (GPT-4) \cite{brown2020language} to generate analogical inspirations (Figure \ref{fig:sketch2design}a).
We leverage prior techniques in chain-of-thought reasoning \cite{wei2022chain} to break down the problem of creating analogies based on the given concept word. 
We take a two-step prompting approach shown in the listings below. 
We first prompt the LLM to detail the design principles for the given subject (e.g., car design).

\begin{lstlisting}[style=prompt]
Describe the key design principles in <subject> design in one short paragraph.
\end{lstlisting}

Such design principles can be useful as context \cite{lewis2020retrieval} to ground the LLM to ideate inspirations that are more suitable for the specific product domain. 
An example intermediary result from this step for the domain of car design could be
% \newpage
\begin{lstlisting}[style=prompt]
Key design principles for car design include aerodynamics exteriors for fuel efficiency and performance... 
\end{lstlisting}

Given the design principles, we then prompt the LLM to generate analogical inspirations.
Our definition of analogy draws on the work of Gentner \cite{gentner1983structure}. This definition involves identifying parallel relations from a source domain to apply to a target domain even when their surface features differ.
We have structured our prompt using this definition of analogies by finding visually concrete objects from three source domains (nature, architecture, fashion) that convey concepts of the target domain (abstract concept).

\begin{lstlisting}[style=prompt]
You are a <subject> designer. The design principles in <subject> design are as follows: <design principles from Step 1>. Brainstorm analogical inspirations for <subject> design to convey a sense of <concept> from one of the following domains: nature, architecture, or fashion. Answer in a bullet-point list of 10 items (item1\nitem2...\nitem3) using visually-concrete objects not adjectives and don't repeat.
\end{lstlisting}

We empirically found that prompting specifically for the domains of nature, architecture, and fashion leads to particularly interesting inspirations. 
Furthermore, these domains are outside of the primary product design domain and are common sources of inspiration for designers.
The LLM then provides single noun-phrases as results (e.g., \texttt{protectiveness} \& $\rightarrow$ \texttt{tortoise, armadillo, armor}).

The user may select a recommended inspiration and continue branching out to explore further inspirations.
For example, selecting tortoise and rerunning the analogy inspiration chain could result in new analogies such as \texttt{tortoise} $\rightarrow$ \texttt{tank, backpack, treasure chest}).
The user may also freely change their concept (e.g., changing \textit{protectiveness} to \textit{freedom}, resulting in a new set of analogical inspirations).
In our current implementation, users cannot return to previous inspirations or explore multiple inspiration branches in parallel. When a user selects an inspiration, it serves as a base to generate another set of inspirations. Furthermore, the generated inspirations remain independent of what the user sketches on the canvas. These features suggest potential areas for future work.

After selecting an analogical design inspiration, the user may create product designs by sketching on a canvas. 
Our conversations with the professional design team revealed that they often start a design with a single silhouette line (Design Goal 1).
In Inkspire, the user may start generating images with as little as a single stroke (Figure \ref{fig:sketch2design}b). Using ControlNet \cite{zhang2023adding} to guide Stable Diffusion \cite{rombach2022high}, we generate a product design guided by the initial stroke. The user may continue adding additional strokes. Each time a stroke is drawn, we generate a new design, making the creation process iterative and implicitly encouraging users to focus on sketching instead of engineering text prompts (i.e., the current paradigm of working with T2I models).

ControlNet does not support per-stroke interaction out-of-the-box as it struggles with incomplete sketches, which is especially problematic during the initial stages of user sketching. Thus, we adapted the ControlNet model with a dynamic guidance scale to enable our desired per-stroke interaction (Equation \ref{eq:guidance}).
The guidance scale is a parameter for controlling how closely the model adheres to user input.
We initialize with a low guidance scale to handle incomplete sketches. We progressively increase the guidance scale as the designer adds more ink to the sketch, pushing the generations to become more sensitive to the user's sketch over time.

\begin{equation}
    G(n) = 7 - 4 \cdot 0.5^{\frac{n}{3}}
    \label{eq:guidance}
\end{equation}
where $G(n)$ is the guidance scale at $n$ number of strokes by the user. The guidance scale starts at 3 when $n$=0 and approaches a maximum of 7 as the sketch becomes more complete ($n$$\approx$10). The decay term $0.5^{\frac{n}{3}}$ creates a sharp logarithmic growth in guidance scale that converges gradually at the maximum value.

By gradually increasing the guidance scale, Inkspire allows the user to more precisely guide the generations as their sketches become more complete and well-defined. We maintain the same initial seed (seed specifies the random noise used to initialize image generations) between generations to maintain \textit{consistency} between iterations and for \textit{near-real-time} generations. The user may click on the "remix" button to change to a different seed and generate more diverse designs that break from the current thread that the designer is exploring. Finally, we remove unnecessary backgrounds from the generated design using a foreground extraction method \cite{qin2020u2} (Figure \ref{fig:sketch2design}c).

\begin{figure*}[tbp]
  \centering
  \includegraphics[width=\textwidth]{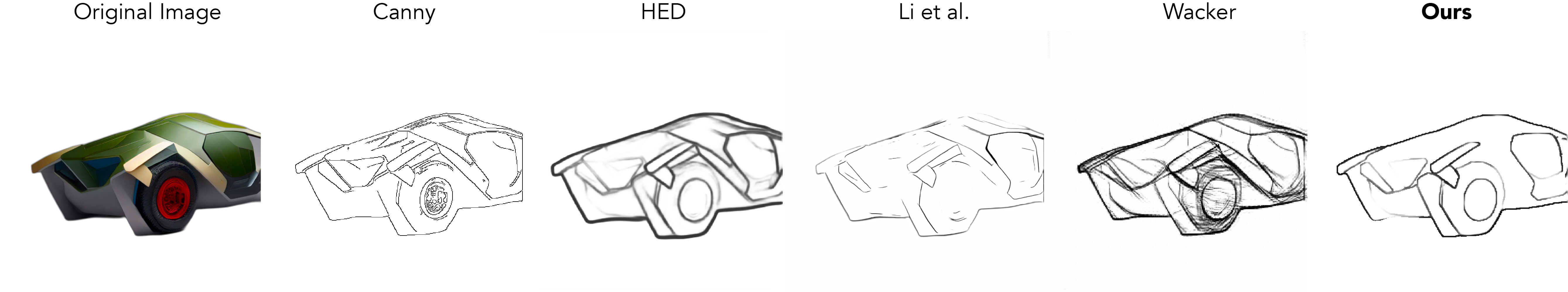}
  \vspace{-1em}
  \caption{Comparison of our Design2Sketch method with potential alternative methods, such as edge detection, manga line extraction models, and models trained explicitly on pairs of sketches and images.}
  \Description{Comparison of our Design2Sketch method with potential alternative methods, such as edge detection, manga line extraction models, and models trained explicitly on pairs of sketches and images.}
  \label{fig:sketchify-comparison}
\end{figure*}

\subsection{Design2Sketch}

\begin{figure*}[tbp]
  \centering
  \includegraphics[width=15cm]{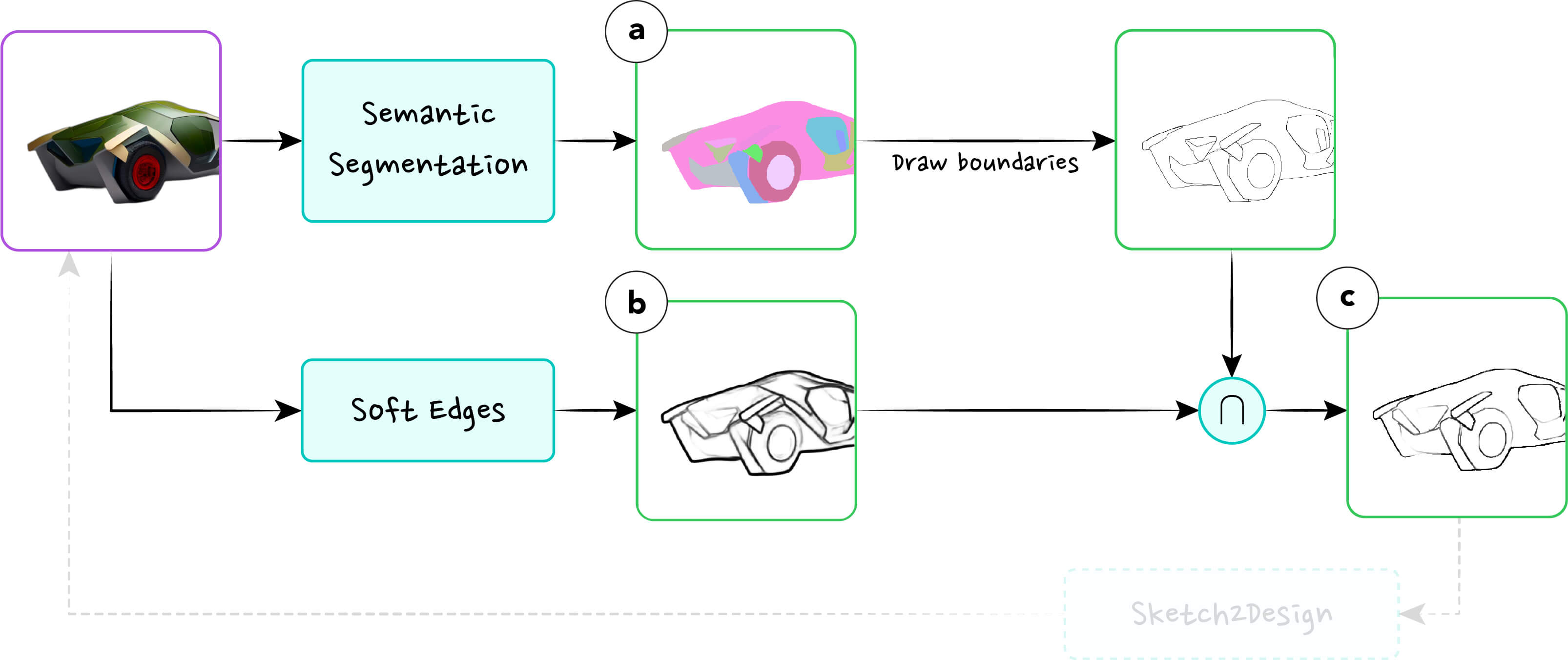}
  \caption{Design2Sketch pipeline, including (a) semantic segmentation, (b) soft edge extraction, and (c) computing an intersection.}
  \Description{Design2Sketch pipeline, including segmentation, soft edge extraction, and computing an intersection.}
  \label{fig:design2sketch}
\end{figure*}

The \textit{Design2Sketch} component helps users build on top of previously generated designs by converting them into \textit{scaffoldings} by abstracting a design into a reduced-fidelity sketch-style with the aim of reducing design fixation on the high-fidelity image (Figure \ref{fig:design2sketch}). The scaffolding appears as a transparent underlay beneath the user's canvas, functioning like tracing paper \cite{tracing} that updates in real-time as they sketch. This enables the user to draw inspiration from aspects of the previously generated designs and also helps them overcome the challenge of starting with a blank canvas \cite{joyce2009blank}, especially during the early stages of sketching. The user can continue iterating through sketching, completing the sketch-to-design-to-sketch feedback loop (Design Goal 3).

While there are many methods for reducing high fidelity images into lower-resolution, we introduce a novel approach for converting designs to sketch scaffolds.
We initially tested existing methods including Canny edge detection \cite{canny1986computational}, HED soft edge extraction \cite{xie2015holistically}, a state-of-the-art method for extracting main lines from manga illustrations \cite{li2017deep}, and a neural network method explicitly trained on pairs of sketches and images. (see Figure \ref{fig:sketchify-comparison}).
We observed that edge extraction methods including Canny edge detection and HED soft edge extraction frequently produce unwanted lines caused by the texture of designs.
Furthermore, we observed that manga line extraction methods, trained primarily on cartoon illustrations, can lead to a loss of key lines or produce broken lines.
Finally, neural networks explicitly trained on pairs of sketches and images on the task of translating images to sketches can create artifacts of excessive sketch stylization, such as shading effects.

In our approach, we combine semantic segmentation and edge extraction.
First, we perform semantic segmentation on the design \cite{kirillov2023segment} to create a segmentation map that color-codes a design into distinct semantic regions (Figure \ref{fig:design2sketch}a). Given the segmentation map, we draw the boundaries between the different regions to create an image of semantic boundary lines. Second, we extract soft edges from the design \cite{xie2015holistically} (Figure \ref{fig:design2sketch}b).
These soft edges include varying thickness and line weight, simulating a sketch-like look, though often with many redundant lines caused by texture.
Finally, we take a pixel-wise intersection between the segmentation map boundary lines from the first step and the extracted soft edges from the second step as the final scaffolding (Figure \ref{fig:design2sketch}c):

\begin{equation}
\textit{Scaffolding} = \textit{Boundary}(\textit{Seg}(D)) \cap \textit{SoftEdge}(D)
\end{equation}
where $D$ is the generated design.
Through this approach, we are able to acheve the best of both worlds -- creating a sketch scaffolding that achieves a natural sketch look while focusing only on the design's key structural lines, filtered through the boundary lines from semantic segmentation step.

\section{User Study}

We conducted a within-subjects study to understand how Inkspire could address designers' pain points in working with T2I models, its potential to be integrated into design workflows, and identify areas for improvement. We compared Inkspire against a baseline condition using a typical ControlNet \cite{zhang2023adding} setup consisting of a prompt box and a sketching canvas, adopting a similar interface layout as Inkspire (see Figure \ref{fig:baseline}). 

\subsection{Participants}
We invited twelve participants (P1-P12, 10 male and 2 female) to participate in a one-hour user study. Among the participants, six are professional designers who perform product design activities daily or weekly (self-rated confidence in product design $\mu$=6.17, $\sigma$=0.75; self-rated confidence in drawing $\mu$=6.00, $\sigma$=1.27; 7-point Likert scale) and six are novices who have moderate drawing experience (self-rated confidence in drawing $\mu$=4.17, $\sigma$=1.33; 7-point Likert scale) but do not actively engage in product design. The participants were recruited through known contacts and Upwork \cite{upwork}, a platform for hiring freelancers. They were not exposed to the Inkspire system or concept prior to the user study. Participants accessed Inkspire and the baseline tool through a web browser.

\subsection{Measures}

For both conditions, we asked participants to complete questionnaires to capture their perspectives on using both Inkspire and ControlNet.
We assess creativity using the Creativity Support Index \cite{cherry2014quantifying}, measuring exploration, inspiration, engagement, expressiveness, tool transparency, and effort/reward tradeoff.
We assess designers sense of the human-AI collaboration using questions from \cite{lawton2023drawing} measuring controllability, communication, harmony, partnership, attribution, and ownership. 
We also asked participants to rate their experiences of sketching by referencing the sketching principles from Bill Buxton's Sketching User Experiences \cite{buxton2010sketching}, measuring how quick and timely, inexpensive and disposable, and loose and abstract sketching with each tool felt.
We asked designers to rate the quality of the final design created with Inkspire and ControlNet and their overall experience satisfaction using each tool.
All questionnaire questions were rated on a 7-point Likert scale (7=highly agree, 1-highly disagree).
We compared the questionnaire measures using parametric paired \textit{t}-tests.
In addition, we log user interaction data, such as when participants draw a new sketch stroke, edit a prompt, and generate a new design. (Figures \ref{fig:log-all}).

\subsection{Procedure}

\subsubsection{Introduction (5 minutes)}

Participants provided informed consent and were given an overview of the study procedures. In addition, we briefly explain how components of the system, such as the analogy generation and sketching and scaffolding interaction, works.

\subsubsection{Design Tasks (45 minutes)}

Participants completed a design task with Inkspire (Figure \ref{fig:teaser}) and another design task with the baseline ControlNet tool (Figure \ref{fig:baseline}). 
The two design tasks are "design a lamp with the theme of serenity" and "design a chair with the theme of fluidity." We counterbalance both the order of the tools and the order of the design tasks. After each condition, the participants completed the questionnaires. 

\subsubsection{Post-Study (10 minutes)}

Participants gave feedback during a short interview as well as through a free response questionnaire on the individual subcomponents of Inkspire, their overall experience of using Inkspire, whether they could see Inkspire being integrated into their design workflow, and areas for improving the tool.
We reviewed these qualitative data to support the quantitative results.

\begin{figure*}[tbp]
  \centering
  \includegraphics[width=\textwidth]{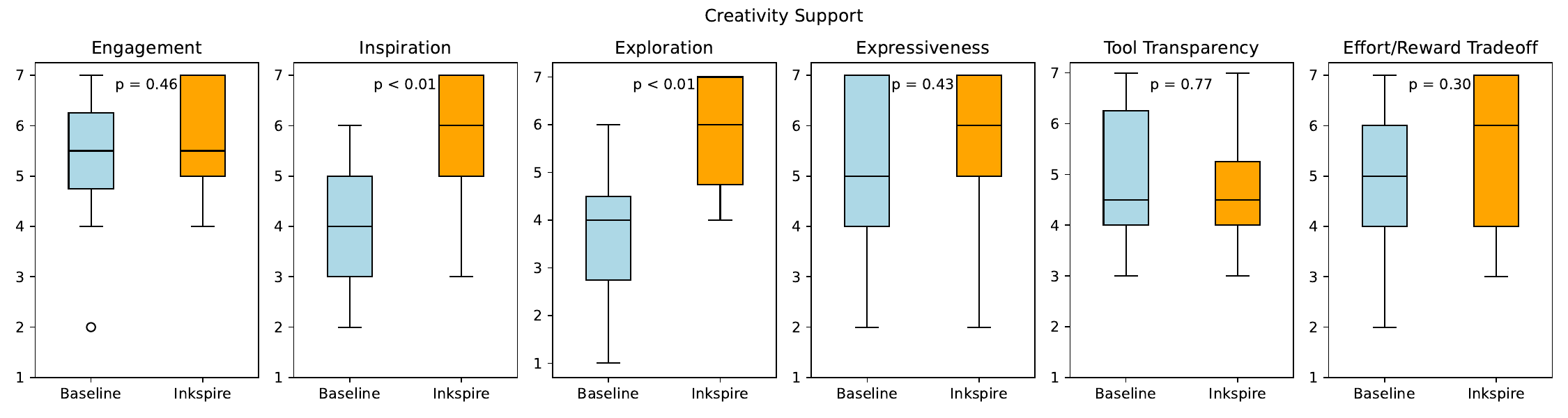}
  \caption{Results on creativity measured with the Creativity Support Index (CSI) \cite{cherry2014quantifying} (7-point Likert scale, higher is better).}
  \Description{Results on creativity measured with CSI (7-point Likert scale, higher is better).}
  \label{fig:csi}
\end{figure*}

\begin{figure*}[tbp]
  \centering
  \includegraphics[width=13.5cm]{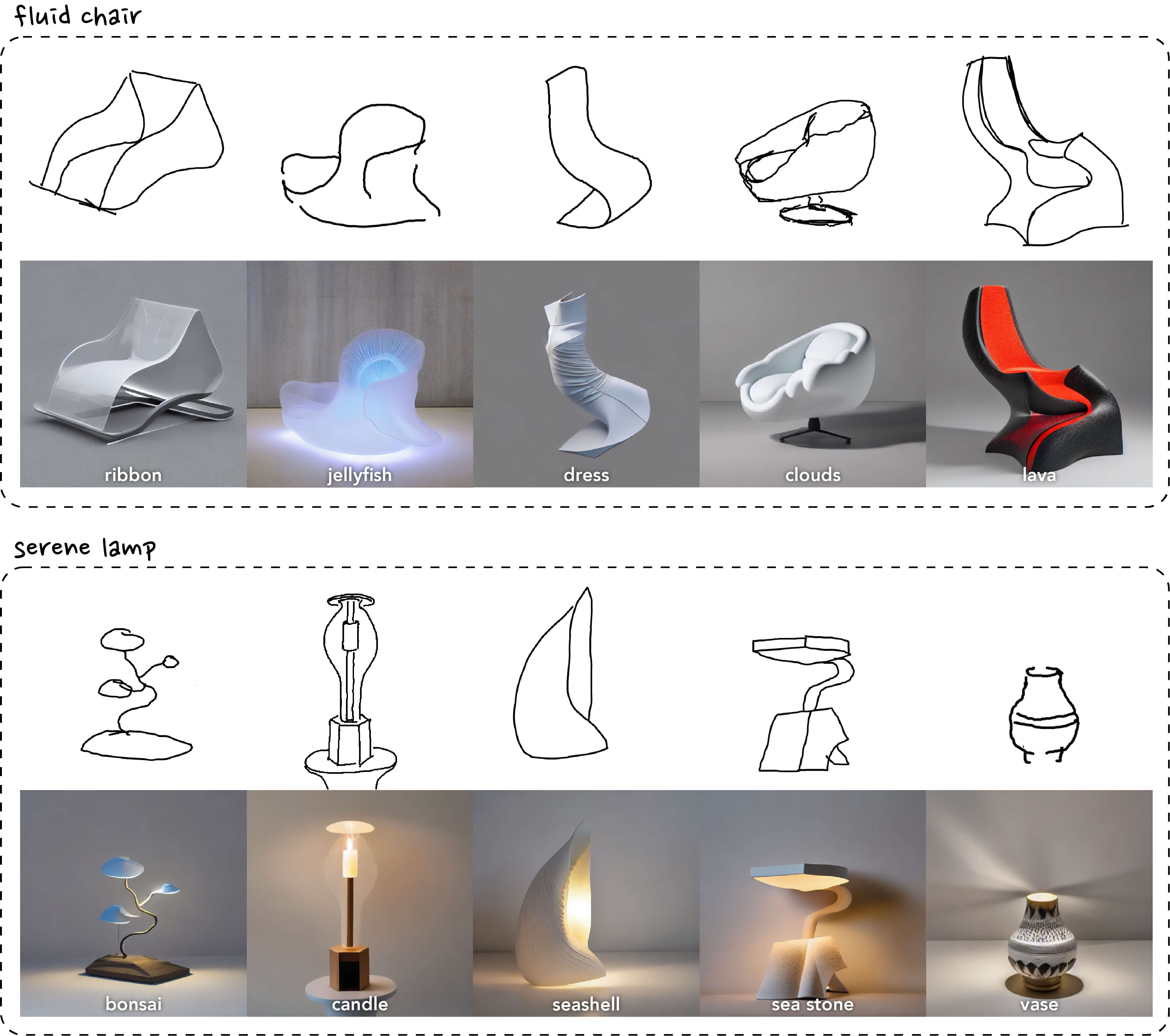}
  \caption{Example designs created by participants using Inkspire for the design tasks of designing a fluid chair and a serene lamp. The final participant-generated sketch is shown on the top, and the generated T2I image is shown on the bottom, along with the selected analogy word chosen by the participant.}
  \Description{}
  \label{fig:gallery}
\end{figure*}

\section{Results}

\subsection{Creativity}
Participants felt that Inkspire improved support for creativity across some attributes of the Creativity Support Index (CSI), shown in Figure \ref{fig:csi}. 
Notably, participants reported significantly higher exploration with Inkspire ($\mu$=5.83, $\sigma$=1.27) as compared to the baseline ($\mu$=3.83, $\sigma$=1.64), ($t$(11)=3.94, $p$<0.01, $r$=0.77, $d_s$=1.13).
Participants also reported significantly higher inspiration ($\mu$=5.92, $\sigma$=1.24) as compared to the baseline $\mu$=4.00, $\sigma$=1.41), ($t$(11)=3.44, $p$<0.01, $r$=0.72, $d_s$=0.99). 

From our interviews, participants noted that the [analogy] inspirations feature is "\textit{helpful while doing design ideations}" (P12) and "\textit{a good tool to brainstorm in the early stage of design}"(P5).
Participants also found Inkspire to effectively support the exploration of multiple ideas, such as a "\textit{variety of forms, styles, patterns, and proportions}"(P5).
P11 explained that they could explore different design directions by manipulating high-level concepts ("\textit{I could just make a basic shape, and change a few keywords and the entire look and feel would change and present me with some great concepts}").
In contrast, we observed that in the baseline condition, many participants focused on sketching extensively on a single idea, a potential sign of the "sunk-cost effect" \cite{arkes1985psychology} (the more time spent in a given direction, the harder it is to move to a different one).

We also note that the other attributes of the CSI, Engagement, Expressiveness, Tool Transparency, and Effort/Reward Tradeoff, were not significantly different between Inkspire and the ControlNet baseline.
While Inkspire did not improve these aspects of creativity over using ControlNet it did not appear to degrade them either.
Overall, we find that Inkspire improved exploration and inspiration, supporting our original design goals and working toward reducing design fixation.

\begin{figure*}[tbp]
  \centering
  \includegraphics[width=\textwidth]{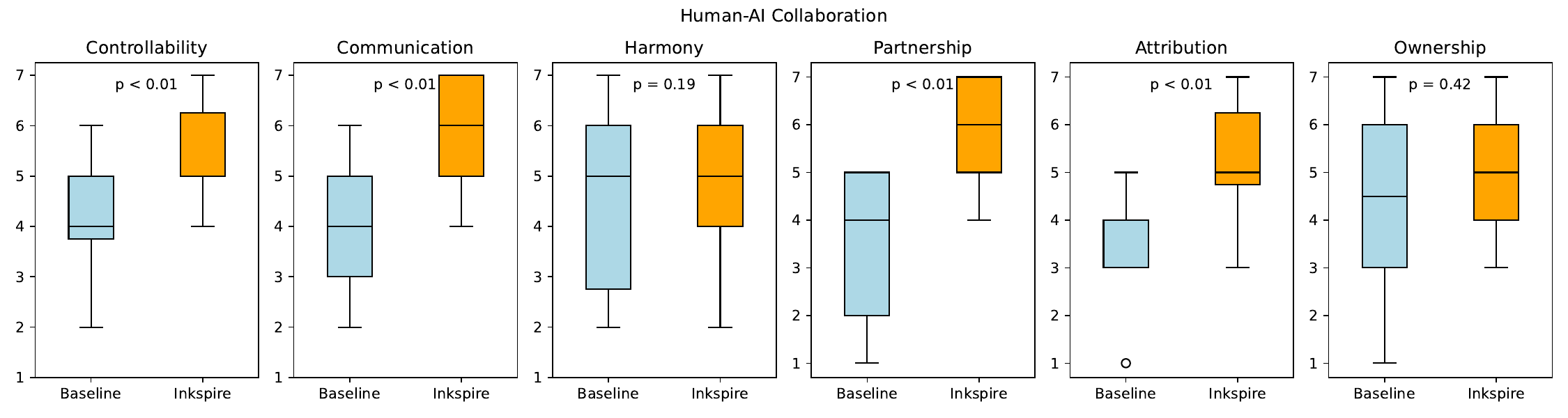}
  \caption{Results on human-AI collaboration measured with Human-Machine Collaboration Questions from \cite{lawton2023drawing} (7-point Likert scale, higher is better).}
  \Description{Results on human-AI collaboration measured with (7-point Likert scale, higher is better).}
  \label{fig:hai}
\end{figure*}

\subsection{Human-AI Collaboration}
\subsubsection{Designer self-ratings of Human-AI Collaboration}

Participants felt that Inkspire improved the experience of collaborating with the AI, across dimensions of Human-Machine Collaboration questions from \cite{lawton2023drawing} rated on 7-point Likert scales, shown in (Figure \ref{fig:hai}).
Participants reported significantly higher controllability when using Inkspire ($\mu$=5.58, $\sigma$=1.00) as compared to the baseline ($\mu$=4.17, $\sigma$=1.19), ($t$(11)=3.56, $p$<0.01, $r$=0.73, $d_s$=1.03).
Participants felt that they had significantly higher communication with Inkspire ($\mu$=5.75, $\sigma$=1.14) as compared to the baseline ($\mu$=3.92, $\sigma$=1.38), ($t$(11)=4.52, $p$<0.01, $r$=0.81, $d_s$=1.31).
In addition, participants rated having a significantly higher degree of partnership with the AI when using Inkspire ($\mu$=5.83, $\sigma$=1.03) as compared to the ControlNet baseline ($\mu$=3.42, $\sigma$=1.56), ($t$(11)=4.57, $p$<0.01, $r$=0.81, $d_s$=1.32).
Lastly, participants rated having significantly more of their own attribution in the designs when using Inkspire ($\mu$=5.25, $\sigma$=1.29) as compared to the baseline ($\mu$=3.67, $\sigma$=1.07),($t$(11)=3.51, $p$<0.01, $r$=0.73, $d_s$=1.01).
Participants did not report significant differences in their feelings of harmony with the AI or ownership over the designs between Inkspire and baseline ControlNet.
% These results may suggest that Inkspire created significant improvements on several aspects of co-creativity (improved controllability, communication, partnership).

\begin{figure*}[tbp]
  \centering
  \includegraphics[width=12cm]{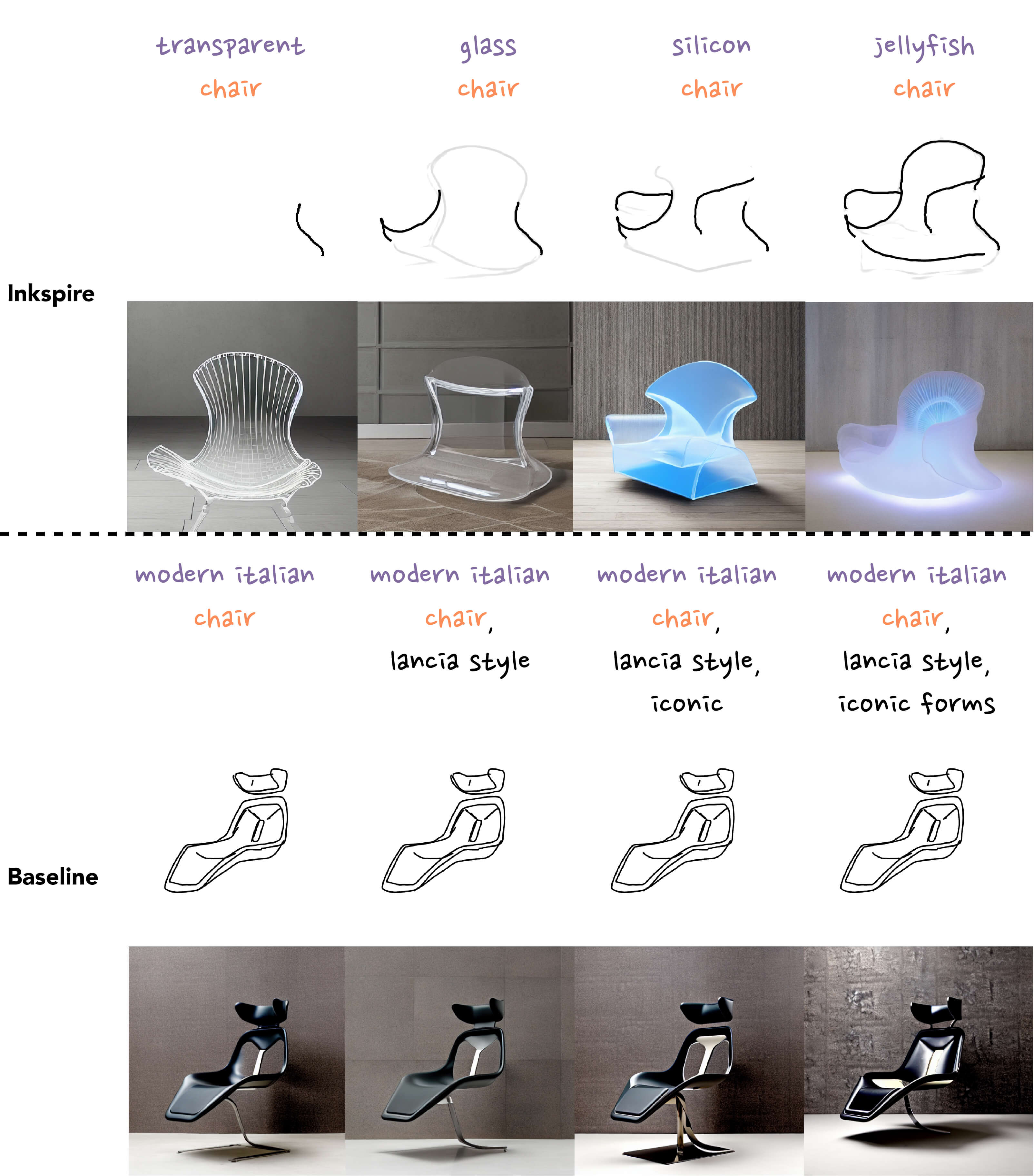}
  \caption{Example user iterations with Inkspire (top section) vs. the baseline condition (bottom section). For each section, the top row shows the prompts, the middle row shows the sketches and scaffolds, and the bottom row shows the generated designs. Using Inkspire, users create diverse designs via analogies. In addition, users start with a single sketch line and continuously build on their sketch, with the guidance of scaffolding. In contrast, using the baseline, users typically draw a full sketch and update their prompt with incremental modifications. This may lead to a higher degree of fixation and a smaller design space explored.}
  \Description{}
  \label{fig:intermediate}
\end{figure*}

From our interviews and open-response questions, many participants commented that they found the scaffolding helpful in being able to \textit{understand the current state of the AI} and plan subsequent sketches ("[The scaffolding] \textit{was useful because it let me know where the current iteration was so I could tell where I'd like to move it next}" (P11)). With scaffolding, participants felt that they could \textit{steer} the direction of the design by building on previous generations ("[scaffolding] \textit{helped with building upon the previous AI-generated design and gave me a direction for what to adjust}" (P6), "\textit{the AI-generated drawing overlay help[ed] me to draw my next line}"(P3), "\textit{having the [scaffolding] as reference helped me [to] combine and remix [old designs] in my [new] sketches }"(P6)).
These observations and self-reported results suggest that Inkspire improves human-AI collaboration by increasing controllability, communication, and partnership.

% This is interesting as the way participants interact with Inkspire is vastly different than how they are used to interact with T2I tools (please see usage logs in Figures \ref{fig:log-inkspire} and \ref{fig:log-baseline}).
% Moreover, participants felt a similar level of ownership over the design outcomes regardless of the tool, but had a significantly higher sense of attribution when using Inkspire.

\subsubsection{Prompting Behavior}
As expected, we observed that participants generally did less prompt engineering when using Inkspire ($\mu$=8.50,  $\sigma$=6.31, number of prompts) than when using the baseline ($\mu$=11.9,  $\sigma$=10.6, number of prompts). 
When using the baseline ControlNet system, participants often relied on manipulating the prompt to change their design whereas they used more diverse analogical inspirations as prompts when using Inkspire.
Analyzing the semantic similarity between user prompts, we observed a much lower semantic similarity in the Inkspire condition ($\mu$=0.51, $\sigma$=0.08) compared to the baseline ($\mu$=0.76, $\sigma$=0.13), measured with BERTScore \cite{zhang2019bertscore}.

Looking at the logs of prompts designers used, we observed that participants using ControlNet often stuck to their original prompt, making small edits to make the prompt more and more detailed (see Figure \ref{fig:intermediate}).
For example, P3 started with the prompt \texttt{serenity lamp} (i.e., explicitly prompting the AI with the abstract design task), and then expanded it with additions like \texttt{serenity lamp with clear glass and black base}, \texttt{serenity lamp with clear glass and black pedestal}, and so on.
This result echoes the prompt fixation results of \cite{Wadinambiarachchi2024effects} and adds further evidence to the challenges designer have in knowing how to prompt \cite{zamfirescu2023johnny}.
Using Inkspire, participants made fewer manual prompt edits and frequently utilized the recommended analogical inspirations to \textit{guide} their prompting direction.
For example, the same participant above (P3) used various analogical inspirations for creating a \textit{fluid} chair such as \texttt{silk}, \texttt{river}, and \texttt{waterfall}.

\subsubsection{Sketching Behavior}
We observed that participants drew fewer total sketch strokes using Inkspire ($\mu$=17.3, $\sigma$=7.40, number of strokes) as compared to using the baseline ($\mu$=59.8, $\sigma$=40.5, number of strokes).
When using Inkspire, participants also had a lower sketching frequency ($\mu$=12.5, $\sigma$=12.5, strokes/min) as compared to the baseline ($\mu$=20.1, $\sigma$=12.6, strokes/min), though this was not significant ($t$(11)=-1.67, $p$=0.12, $r$=0.45, $d_s$=0.48). 
Participants also spent more time between strokes when using Inkspire ($\mu$=13.8s, $\sigma$=12.1s) as compared to the baseline ($\mu$=4.44s, $\sigma$=2.94s), ($t$(11)=2.66, $p$=0.02, $r$=0.63, $d_s$=0.77).

\begin{figure}[tbp]
  \centering
  \includegraphics[width=8.5cm]{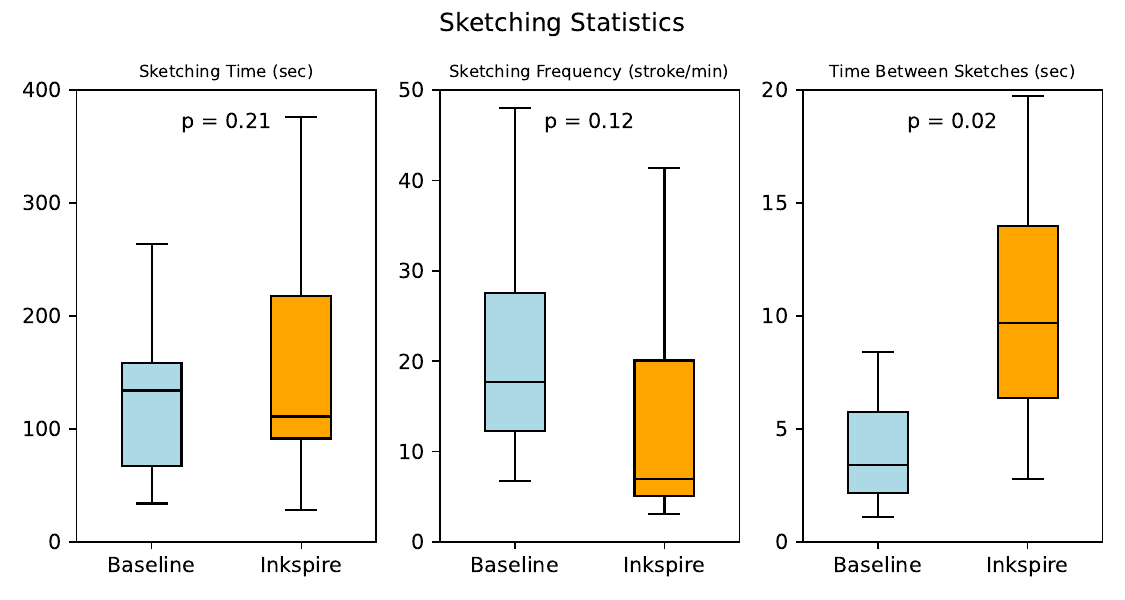}
  \caption{Results on sketching statistics, including total sketching time, sketching frequency, and time between sketches.}
  \Description{Results on sketching statistics, including total sketching time, sketching frequency, and time between sketches.}
  \label{fig:sketching-stats}
\end{figure}

Despite participants sketching less and taking longer between adding more ink to their drawing, participants rated that sketch strokes felt significantly more inexpensive when using Inkspire ($\mu$=6.08, $\sigma$=0.90, 7-point Likert scale, higher is better, based on Buxton's sketching principles \cite{buxton2010sketching}) as compared to using the baseline ($\mu$=3.75, $\sigma$=1.60, 7-point Likert scale, higher is better), ($t$(11)=5.01, $p$<0.01, $r$=0.83, $d_s$=1.45).
Participants also rated that the sketching was more abstract when using Inkspire ($\mu$=5.75, $\sigma$=1.14, 7-point Likert scale, higher is better, based on Buxton's sketching principles \cite{buxton2010sketching}) as compared to the baseline ($\mu$=3.83, $\sigma$=1.47, 7-point Likert scale, higher is better), ($t$(11)=4.24, $p$<0.01, $r$=0.79, $d_s$=1.23).
Overall, Inkspire appears to have provided designers with an interface for sketching ideas collaboratively with AI that improves over the more stilted sketching and image generation experience of other text-to-image systems.

\begin{figure*}[tbp]
  \centering
  \includegraphics[width=12cm]{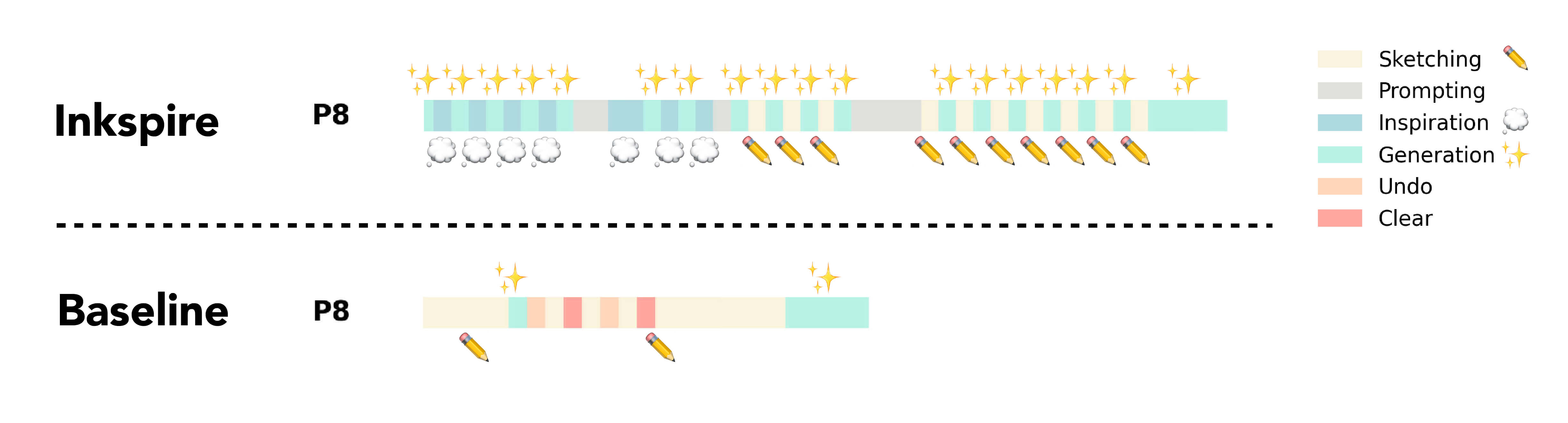}
  \caption{Example user interaction log. We observe that, using Inkspire, the user started by ideating several analogical inspirations (thought bubble). Next, they performed sketching in a highly iterative manner (pencil+sparkles). Overall, the user generated a significant amount of new designs (sparkles). In contrast, using the baseline condition, the user sketched in large stretches with infrequent new design generations. We see that this pattern generally holds true across participants (please see Figure \ref{fig:log-all} for data on all participants).}
  \Description{Example log.}
  \label{fig:log-example}
\end{figure*}

\subsubsection{Analogy Inspirations}
We observed that participants generally explored multiple analogical inspirations ($\mu$=4.58, $\sigma$=2.43) distributed across the three categories (Nature $\mu$=2.08, $\sigma$=1.56; Architecture $\mu$=1.58, $\sigma$=1.3; Fashion $\mu$=0.917, $\sigma$=0.716) (Table \ref{table:analogy-categories}). Architecture was the most common final choice ($n$=6 instances across participants), followed by Nature ($n$=4) and Fashion ($n$=2), though participants explored Nature-based inspirations more frequently on average.
We also observed that participants frequently switched between analogy categories.
For instance, P12 quickly switched between all three categories.
Among the switches, we observed that Nature→Architecture ($n$=6) and Nature→Fashion ($n$=5) were common transitions. This may suggest that Nature can serve as a common "bridge" category.
Many participants' final category deviated from their initial category after exploration. For instance, P2 started with Nature and selected Architecture after exploring 14 different inspirations across all categories.
Overall, participants drew inspiration from all categories, with Nature playing a central role, though each category proved to be a valuable source for inspiration.

\subsubsection{Overall Usage Patterns}
By analyzing our logged interaction data, we observed common usage patterns for Inkspire vs. the baseline ControlNet workflow (please see Figure \ref{fig:log-example} for an illustrative example and Figure \ref{fig:log-all} for the full data).
First, participants using Inkspire often start the initial ideation phase by experimenting with many analogical inspirations to explore the design space (shown as many thought bubbles during the beginning, Figure \ref{fig:log-example}). This early divergent exploration behavior may suggest that designers were able to consider a wide range of inspirational ideas at the start of their process. Subsequently, participants using Inkspire engaged in a highly iterative manner of sketching and interacted with scaffolding, often drawing one or a small number of strokes, seeing a generation and scaffolding, then pausing a bit to consider where to move next (shown as many pencils interleaved with sparkles, Figure \ref{fig:log-example}).
% Participants used the scaffolding underlay extensively, especially during the early stages while the canvas was more empty. 
In contrast, when participants used the baseline system, they often did a complete full sketch (shown as large stretches of a single pencil, Figure \ref{fig:log-example}) before ever hitting the generate button.
Designers made more "Undo" actions, removing strokes that they had made (as shown in orange blocks). 
Overall, participants generally created more new designs with Inkspire than with the baseline condition (shown as significantly more sparkles in Inkspire than the baseline, Figure \ref{fig:log-example}).
Together, these patterns show how when participants used ControlNet, they focused more on crafting a whole sketch then handing it off to the generative AI rather than working collaboratively back and forth with the AI.

\subsection{Final Design Quality}
Participants rated their final creations as having higher design quality when using Inkspire ($\mu$=5.92, $\sigma$=1.08) as compared to the baseline ($\mu$=4.58, $\sigma$=1.93) on a 7-point Likert scale, shown in Figure \ref{fig:overall} (left). Though this difference of ratings for design quality are not statistically significant, ratings showed less spread than when using the baseline ($t$(11)=2.11, $p$=0.06, $r$=0.54, $d_s$=0.61).

In Figure \ref{fig:gallery}, we show several example creations from participants.
While the quality of output in general appears relatively high, we do observe that the final results produced with Inkspire do appear to be more diverse than the results produced with ControlNet.
Looking at the trend of designs created during a user session, we see that the thread of designs created using ControlNet often do not change tracks, whereas those done with Inkspire show more diversity of conceptual exploration. An example showing chair designs from two different users illustrates this in Figure \ref{fig:intermediate}.
The participant using Inkspire (top) shows a wide exploration of concepts for a \texttt{transparent} chair including \texttt{glass}, \texttt{silicon}, and \texttt{jellyfish}.
Alternatively, a participant using ControlNet creates a \texttt{modern Italian} chair with a high quality initial sketch and iterations on the prompt from \texttt{lancia style} to \texttt{lancia style, iconic} to \texttt{lancia style, iconic forms}.
While both final result are generally of high quality, the final output from the participant using Inkspire is derived from a more exploratory process over design analogies over the more fixated design process from the designer working with ControlNet.

\subsection{Designer Ratings of Usage Experience Satisfaction}
\begin{figure}[t]
  \centering
  \includegraphics[width=6cm]{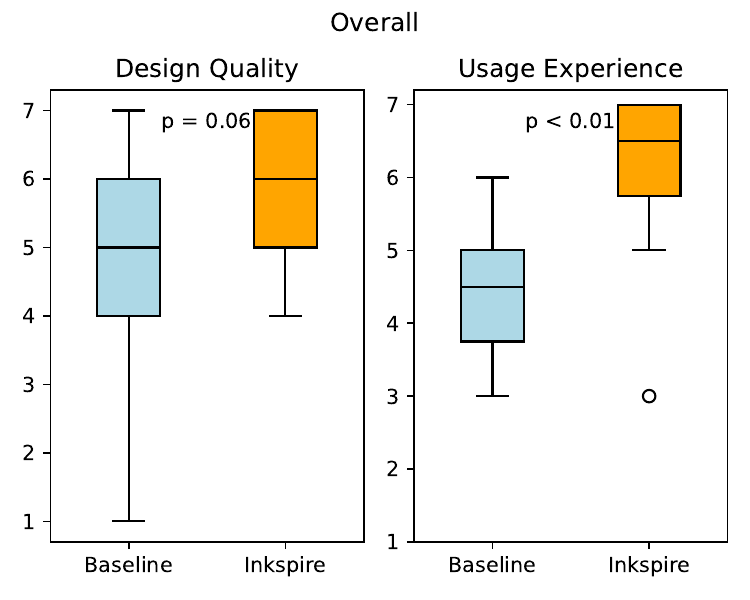}
  \caption{Results on design quality (left) and usage experience (right) (7-point Likert scale, higher is better).}
  \Description{Results on design quality (left) and usage experience (right) (7-point Likert scale, higher is better).}
  \label{fig:overall}
\end{figure}

Participants rated significantly higher satisfaction with their usage experience when using Inkspire ($\mu$=6.08, $\sigma$=1.24) as compared to the baseline ($\mu$=4.33, $\sigma$=0.98) when rated on a 7-point Likert scale ($t$(11)=3.54, $p$<0.01, $r$=0.73, $d_s$=1.02), shown in Figure \ref{fig:overall} (right).

Based on the interaction logs, Inkspire may improve participants' experiences of sketching with AI by supporting low cost experimentation and helping participants draw abstract sketches that focus on the big picture rather than being bogged down by small details.
In the baseline, we observed that participants drew a number of large strokes in succession \textit{before} generating anything, with the majority being "filler" lines (Figure \ref{fig:log-all}). 
This generally led to designers spending more time between trial-and-error attempts to achieve their intended result, possibly reducing their satisfaction when using ControlNet.

% \begin{figure}[tbp]
%   \centering
%   \includegraphics[width=8.5cm]{figures/Prompting_and_Inspiration_Usage.pdf}
%   \caption{}
%   \Description{}
%   \label{fig:prompting}
% \end{figure}

% \begin{callout}
% \justifying
% \emoji{light-bulb} Inkspire supports \textit{initial ideation} (e.g., via seed inspirations) as well as the \textit{expansion into alternative ideas} (e.g., via easy mechanisms to remix designs) helping overcome previously found issues with people's limited prompting abilities leading to design fixation \cite{Wadinambiarachchi2024effects}.
% \end{callout}

% \begin{callout}
% \justifying
% \emoji{light-bulb} Inkspire supports \textit{bidirectional communication} between the user and the AI -- the user first \textit{understands} the current state of the AI in order to \textit{guide} the AI. 

% \end{callout}

\section{Discussion}

We introduced Inkspire, a prototype system to explore new ways for designers to leverage generative AI while avoiding design fixation and embedded into their existing practice of sketching new concepts. 
We focused on a challenge identified in design research literature that exploring a wider range of ideas can lead to improved design outcomes, though designers often find it difficult to naturally engage in diverse design exploration \cite{dow2010parallel, kocienda2018creative}.
Inkspire helped designers explore a more diverse design space by recommending them with diverse analogical anchors based on their initial design concept (e.g., turtle, shield, or bunker for embodying the concept of "protective").
These analogical concepts were used as input along with the sketch from the designer to generate new images. The designer could then continue to iterate on the designs collaboratively with the AI by leveraging stroke-by-stroke generation and building on low-resolution sketch scaffolds of high fidelity renders.
Through a user study, we found that participants using Inkspire explored a wider design space and qualitatively changed their ideation process to be more conceptually iterative and collaborative with the AI. In contrast, participants in a control condition, in which they used a state-of-the-art ControlNet, primarily focused their efforts on crafting full sketches and making small changes to prompt inputs before having the AI render an image.

A core challenge we grappled with in this work was helping designers to leverage generative AI while avoiding fixation on the outputs of the AI.
We believed this to be especially likely in the context of creative design and sketching because of several factors derived from our formative conversations with professional designers and prior research: text prompting leading to what designers called "unnatural" interaction,
text-to-image models struggling with generating inspiring designs from abstract concepts,
both often due to designer's limited abilities to prompt well \cite{davis2024fashioning,Wadinambiarachchi2024effects,zamfirescu2023johnny,zhang2023generative}
and the AI generating outputs that often look overly "finished", a well-known cause of design fixation \cite{cardoso2009design,cheng2014new,Wadinambiarachchi2024effects}.

Our results suggest a fundamental issue with the structure of prompt- or sketch-focused generative image interfaces, which is that current controls through prompts or sketches can engender an affordance of needing to feel relatively complete in order to convey the designer's intention.
The results from when our participants used ControlNet are similar to results from Wadinambiarachchi et al. \cite{Wadinambiarachchi2024effects} where designers focus most of their effort on minor iteration of prompts.
Needing to provide a rich enough prompt or a full enough sketch may explain why our participants felt less control over the ideation and less communication and partnership with the AI model.

\begin{figure*}[tbp]
  \centering
  \includegraphics[width=11cm]{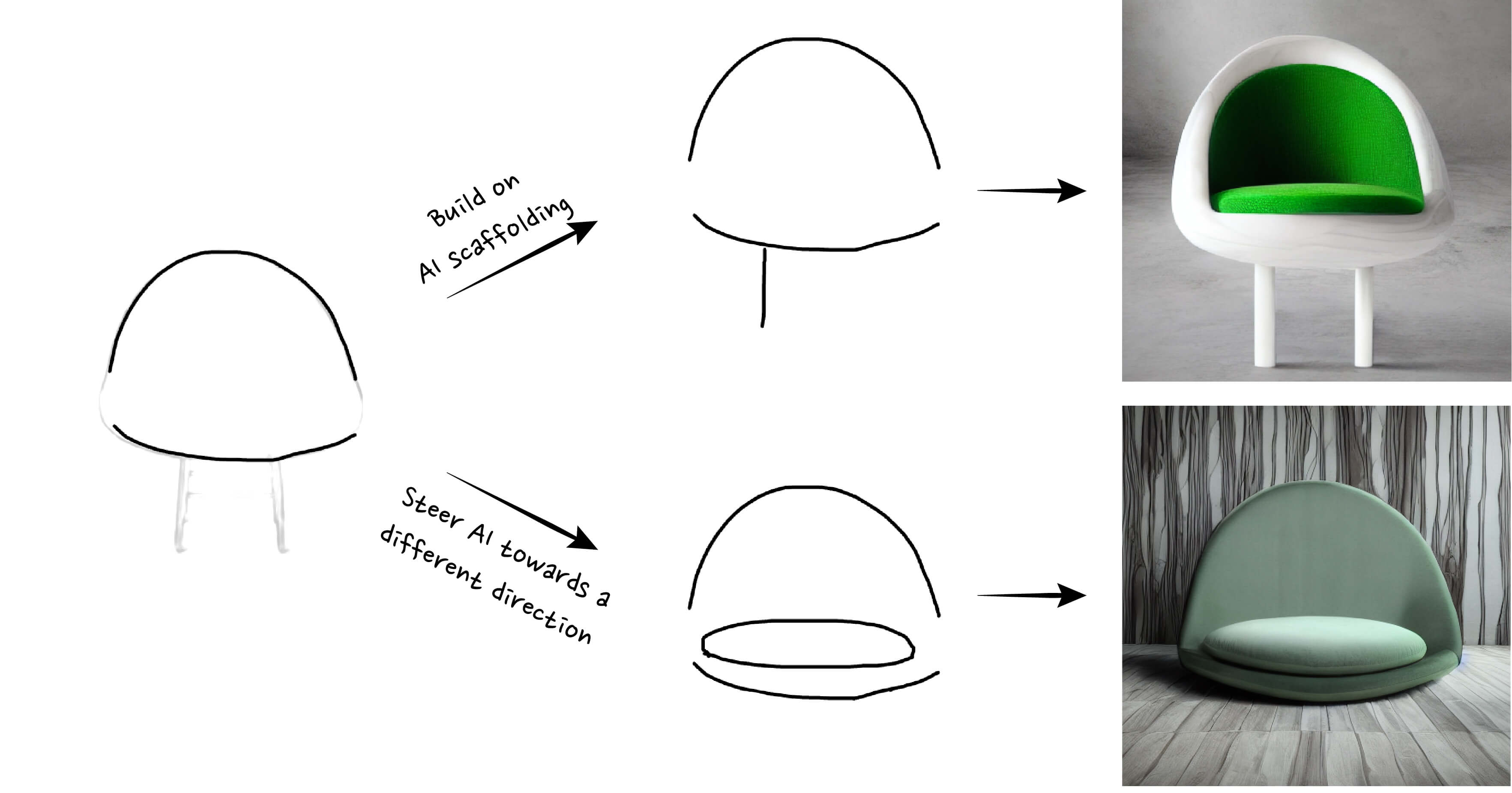}
  \caption{Given a scaffold, the user may choose to build on it (by tracing it) or steer the AI towards a different direction.}
  \Description{}
  \label{fig:branching}
\end{figure*}

In contrast, with Inkspire the design problem prompt and analogical keyword concepts provide a scaffolding in the design space allowing a single stroke to have expressive meaning while also generating an output that is relevant to the user's goals.
Relatedly, our work also explores ways for designers to have a shared mental model of what they are trying to accomplish versus what the AI is trying to do. For example, an interesting point we noticed from our participants was that they found the rough sketches inferred from the AI generation to be useful as control points that showed what the AI was doing and how the user could emphasize or change that, for example by either drawing over or changing a stroke on the roughed sketch, respectively (please see Figure \ref{fig:branching}). In this way the rough sketch serves not only as a scaffold for the designer's iterative sketching but as a communication tool and shared mental model with a generative AI partner.
The continual updating of both sketch and analogical concept ultimately leads to a more iterative and collaborative process where the designer works in turns with the AI.
Other generative AI tools could potentially build on this pattern of rapid turning taking to help promote more co-creativity for their creative task.

Overall, the interactions we designed into Inkspire appear to have worked in helping our participants avoid design fixation.
As commented on by P6 "\textit{the AI was helpful in providing alternative designs I did not consider previously, especially by adopting a new semantic understanding of the sketch that differed from my initial intention}."
However, even though working with Inkspire pushed the designers into a space they had not originally considered and even resulted in them sketching less in total, they found that we also succeeded in overcoming issues where designers feel less agency when working with generative AI.
That the participants felt they had more agency and more ownership of the final design outputs suggests the analogies, stroke-based interaction pattern, and scaffolds led to more fluid interaction through a design space.
Again quoting from P6, Inkspire had "\textit{an interesting workflow that I think helps make the tool disappear more in comparison to a prompt-based drawing tool. It allowed me to focus more on sketching than prompt engineering.}"

While our overall results show that Inkspire achieved our design goals, there are some potential issues with Inkspire that could arise and could be considered in future co-creative systems.
First, although we see more evidence of designers moving to more distinct ideas throughout a session, they still work along a single thread.
Prior research has shown that designers often prototype ideas in parallel and that such parallel exploration can lead to better design outcomes \cite{dow2010parallel}.
One potential way to enable parallel explorations in Inkspire is to generate numerous analogies for a concept and generate an image for each during each sketch step.
Second, as Inkspire was primarily designed for early explorations, it may not serve a designer as well during later stages of the design when they have a clear idea in mind that they are trying to render. 
In this case, a designer may prefer using a tool such as ControlNet where they can focus on the details they want before handing it to an AI. Lee et al. similarly proposed an adaptive multimodal T2I system that initially supports early ideation with a prompt-guided (e.g., protective chair) and sketch-supported (e.g., simple sketch) system, which gradually evolves into sketch-guided (e.g., detailed sketch) and prompt-supported (e.g., "more curved back") system to help refine the concept in later stages of the creative process \cite{lee2024the}. This suggests potential opportunities for generative systems to provide controls for how much turn taking the AI should aim to have with the designer.
Third, Inkspire and other generative AI approaches may help individual designers break their fixation on specific ideas, however, if common generative models are used, it may promote collective design fixation, where everyone's images begin to look the same as they have a common source \cite{anderson2024homogenizing}.
This being said, Inkspire may work to counteract such collective fixation by promoting increased designer engagement by focusing more on suggestions of designs though lower-resolution sketch underlays rather than pixel-level in-painting.
Future systems could explore other techniques for avoiding design fixation, such as showing partial photographs \cite{cheng2014new} instead of sketch underlays. 
Other ideas for reducing fidelity such as blurring or filtering could also promote designers to fill in details on their own, potentially leading to designs less similar to common output from AI models.

\label{section:limitations}
\subsection{Future Improvements to Inkspire}
There are several avenues for future work on improving Inkspire collected from participants' comments.
First, we could improve the sketching canvas with more advanced sketching features such as line weight control and shading tools, which could allow users to sketch with greater detail. 
Second, participants suggested adding more fine-grained region-based control to the Sketch2Design pipeline. 
For example, a user could specify different material types for different regions of the sketch. 
Third, we could extend the analogical panel to include the capability to go back to a previous inspiration and perform multiple branches of inspiration explorations in parallel. This could help users compare vastly different design directions.
Fourth, we currently adopt a two-step prompting mechanism for generating analogical inspirations: defining the design principles for the target domain, then generating visually-concrete object inspirations for the target domain using nature, architecture, and fashion as source domains.
Future work could explore more complex prompting structures, such as traversing hierarchical tree structures \cite{kang2023biospark}, that might lead to better analogical inspirations.
\rev{Fifth, Inkspire contains several features that could all affect a user's behavior, such as automatic per-stroke generation and sketch scaffolding.
For future work, we could conduct additional ablation studies. For example, adding dynamic guidance scale to enable automatic per-stroke generation for the baseline condition or hiding sketch scaffolding for Inkspire.}
Sixth, we adopted a within-subjects design with a relatively small sample size ($n$=12), which could be improved through larger-scale studies.
Seventh, participants suggested the possibility of adding additional guidance for the AI, such as mechanical and material constraints, though such a feature may require more research into how to better connect analogically created T2I design concepts with their feasibility to be manufactured.
Finally, in this paper, we introduce a sketch-generation-scaffold interaction with generated analogies for abstract concepts through the application of product sketching. We think the sketching strategy could generalize to other forms of drawing, though the analogies might be better suited for product design.
For example, very recent work has explored a similar strategy but using an image underlay instead of sketch scaffolding \cite{sarukkai2024block}.
Future work could explore extending the system to additional design domains and incorporating background elements into designs for mockup or decoration.

% This could enable designers to bring their designs closer to the manufacturing stage.
% // Limitations of serial, analogies
% // How does this fit into the future of sketching/CAD tools

\section{Conclusion}
This research identifies the challenges designers face when using Text-to-Image (T2I) models in design and proposes a set of design guidelines for creating new interfaces for interacting with T2I systems. The research prototype, Inkspire, supports designers in prototyping product design concepts with analogical inspirations and through a complete sketch-to-design-to-sketch feedback loop.
Feedback from designers using Inkspire suggests that it could encourage more inspiration and exploration of novel design ideas while allowing designers to focus on iterative sketching collaboratively with the T2I model rather than prompt engineering.
These analogy and sketch scaffolding features allowed designers to generate initial ideas for a design concept from a single abstract concept and single stroke of the pen to expand them into multiple alternatives. 
The iterative interaction and turn-taking between the designer and the AI helped the designer guide the models toward novel design intentions with the potential to overcome design fixation.
We are interested in extending Inkspire with more capabilities and hope that this research can help inform future research on sketch-driven tools for co-creating with Generative AI models.

% Inkspire lets them interaction in a more iterative and reflective way than when using ControlNet.
% this could lead to future systems that afford better fluidity in design work and speak to a kind of interaction that leads to the designers having more control over the generation.
% Analogies can offer a broader design space and can potentially lead to more novel and differentiated designs.
% moreover, we found that we did not negatively impact people's feeling of harmony with the AI system, suggesting that the designers felt comfortable working with the system.

% It also facilitated a convergent refinement process by providing structure for designers to iterate on AI-generated designers. 


\begin{acks}
This work was supported by funding from the Toyota Research Institute and the Office of Naval Research.
We would like to thank Matthew Klenk for valuable conversations and guidance on the research.
We would like to thank friends in the Augmented Design Capability Studio for providing valuable feedback on iterations of our system and paper write-up.
\end{acks}

\bibliographystyle{ACM-Reference-Format}
% \balance
\bibliography{base}
\balance


\appendix
\onecolumn
% \section{Usage Logs of Inkspire and the baseline condition}
% \label{appendix:logs}

\begin{figure*}[tbp]
  \centering
  \includegraphics[width=\textwidth]{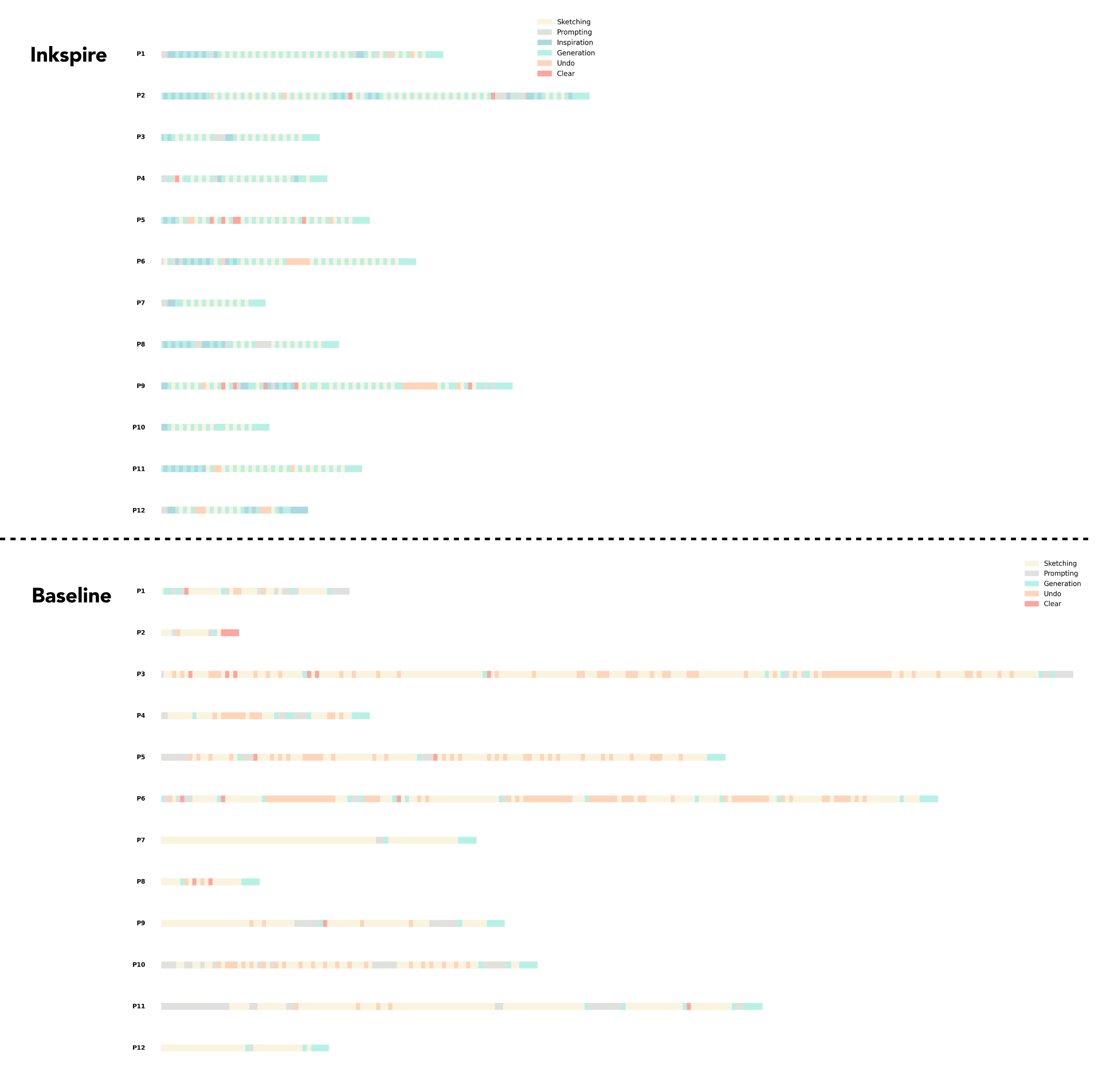}
  \caption{Complete participant usage logs when using Inkspire vs. the baseline. Please see Figure \ref{fig:log-example} for an annotated illustrative example.}
  \Description{Complete participant usage logs when using Inkspire vs. the baseline.}
  \label{fig:log-all}
\end{figure*}

\begin{figure*}[tbp]
  \centering
  \includegraphics[width=\textwidth]{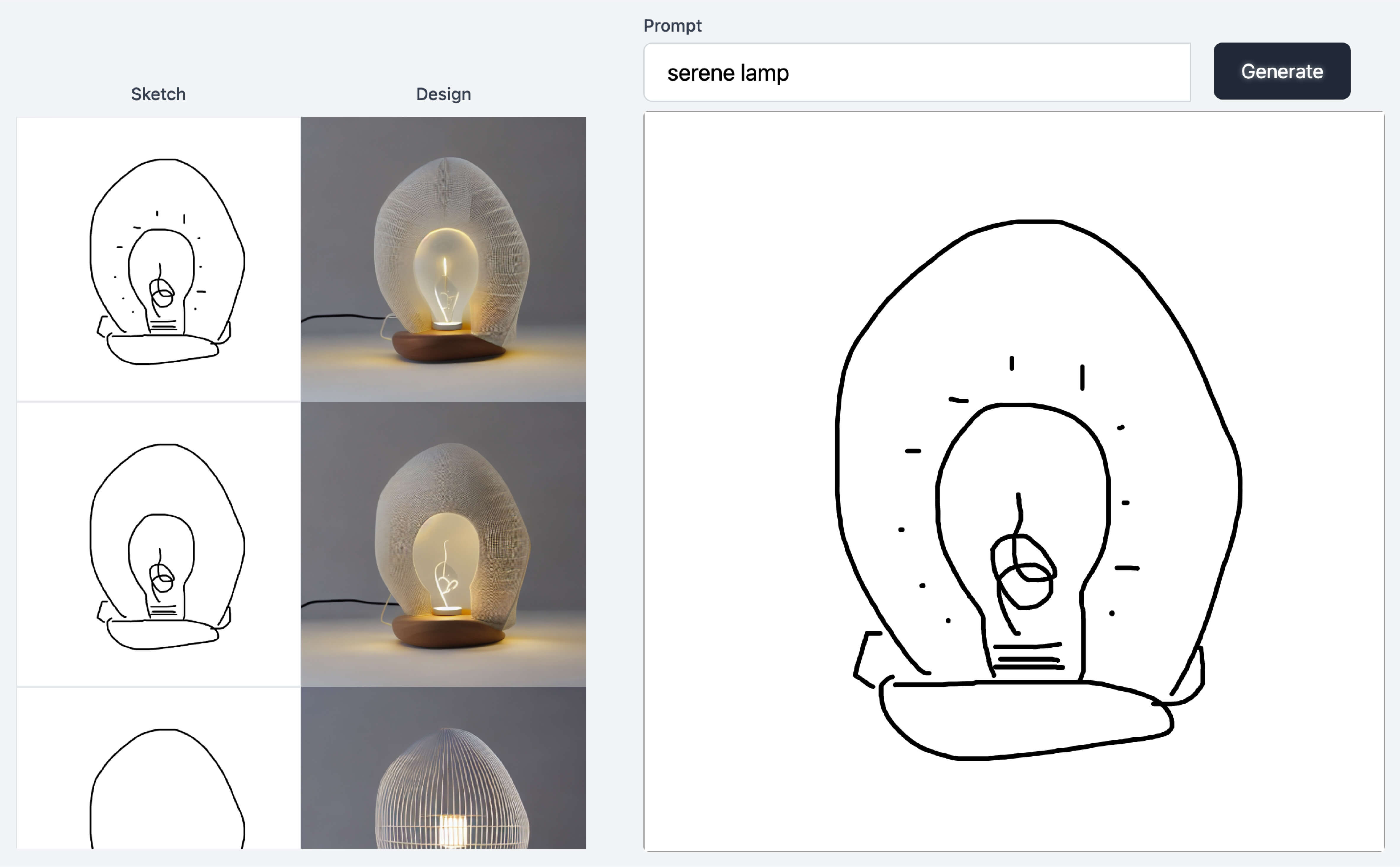}
  \caption{The baseline interface adopts a similar layout as Inkspire, but without analogical inspirations, scaffolding sketch underlays, and stroke-by-stroke generations. The participant would typically manually specify a prompt and draw a complete sketch for the T2I model -- a standard workflow of how designers work with T2I models with additional sketch control \cite{zhang2023adding}.}
  \Description{The baseline interface adopts a similar layout as Inkspire, but without analogical inspirations, scaffolding sketch underlays, and stroke-by-stroke generations. The participant would typically manually specify a prompt and draw a complete sketch for the T2I model -- a standard workflow of how designers work with T2I models with additional sketch control.}
  \label{fig:baseline}
\end{figure*}

\begin{table*}[t]
\small
\begin{tabular}{lp{0.45\textwidth}ccccl}
\toprule
\multirow{2}{*}{ID} & \multirow{2}{*}{Analogies} & \multicolumn{4}{c}{Category Count} & Final\\ 
\cmidrule(lr){3-6}
& & Total & Nature & Architecture & Fashion & Category\\
\midrule
P1 & running cheetah, pleated dress, catwalk, skyscraper, spiral staircase, pleated dress, jellyfish & 6 & 2 & 2 & 2 & Nature \\
P2 & waves, waterfall, silk train, flowing dress, flowing dress, waves, running river, running river, waves, willow tree, infinity pool, concrete, silhouettes of brutalist buildings, rubble & 10 & 4 & 4 & 2 & Architecture \\
P3 & silk, river, clouds, waterfall & 4 & 3 & 0 & 1 & Nature \\
P4 & zen garden, moonlight, candle, minimalist church & 4 & 1 & 2 & 1 & Architecture \\
P5 & bamboo, bonsai, japanese tea house, zen garden & 4 & 2 & 2 & 0 & Architecture \\
P6 & bay window, japanese tea house, zen garden, lotus flower, zen garden, japanese tea house & 4 & 1 & 3 & 0 & Architecture \\
P7 & lotus & 1 & 1 & 0 & 0 & Nature \\
P8 & bamboo grove, moonlight, roman column, snow peak, spring creek, bamboo, zen garden, moonlight & 7 & 5 & 2 & 0 & Nature \\
P9 & waterfall, pawprint, seashell, sea stone, slinky, ribbon & 6 & 4 & 0 & 2 & Fashion \\
P10 & flowing gown & 1 & 0 & 0 & 1 & Fashion \\
P11 & floating bridge, lava, silk scarf, winding staircase & 4 & 1 & 2 & 1 & Architecture \\
P12 & gaudi's architecture, ribbon, waterfall, running track curve & 4 & 1 & 2 & 1 & Architecture \\
\bottomrule
\end{tabular}
\caption{Analogical inspirations explored by participants and statistics on inspiration categories.}
\label{table:analogy-categories}
\end{table*}

% \section{Example user creations}
% \label{appendix:gallery}

\end{document}